\documentstyle[]{mn}

\newif\ifAMStwofonts
 
\ifoldfss
  \newcommand{\rmn}[1] {{\rm #1}}

  \ifCUPmtlplainloaded \else
    \NewTextAlphabet{textbfit} {cmbxti10} {}
    \NewTextAlphabet{textbfss} {cmssbx10} {}
    \NewMathAlphabet{mathbfit} {cmbxti10} {} 
    \NewMathAlphabet{mathbfss} {cmssbx10} {} 
  \fi
  \ifAMStwofonts
    \ifCUPmtlplainloaded \else
      \NewSymbolFont{upmath} {eurm10}
      \NewSymbolFont{AMSa} {msam10}
      \NewMathSymbol{\upi}     {0}{upmath}{19}
      \NewMathSymbol{\umu}     {0}{upmath}{16}
      \NewMathSymbol{\upartial}{0}{upmath}{40}
      \NewMathSymbol{\leqslant}{3}{AMSa}{36}
      \NewMathSymbol{\geqslant}{3}{AMSa}{3E}

      \let\leq=\leqslant 
       
    \fi
  \fi
\fi 
 
\ifnfssone
  \newmathalphabet{\mathit}
  \addtoversion{normal}{\mathit}{cmr}{m}{it}
  \addtoversion{bold}{\mathit}{cmr}{bx}{it}
  \newcommand{\rmn}[1] {\mathrm{#1}}

  \newmathalphabet{\mathbfit} 
  \addtoversion{normal}{\mathbfit}{cmr}{bx}{it}
  \addtoversion{bold}{\mathbfit}{cmr}{bx}{it}
  \newmathalphabet{\mathbfss} 
  \addtoversion{normal}{\mathbfss}{cmss}{bx}{n}
  \addtoversion{bold}{\mathbfss}{cmss}{bx}{n}
  \ifAMStwofonts
    \ifCUPmtlplainloaded \else
      \UseAMStwoboldmath
      \makeatletter
      \new@mathgroup\upmath@group
      \define@mathgroup\mv@normal\upmath@group{eur}{m}{n}
      \define@mathgroup\mv@bold\upmath@group{eur}{b}{n}
      \edef\UPM{\hexnumber\upmath@group}
      \new@mathgroup\amsa@group
      \define@mathgroup\mv@normal\amsa@group{msa}{m}{n}
      \define@mathgroup\mv@bold\amsa@group{msa}{m}{n}
      \edef\AMSa{\hexnumber\amsa@group}
      \makeatother
      \mathchardef\upi="0\UPM19
      \mathchardef\umu="0\UPM16
      \mathchardef\upartial="0\UPM40
      \mathchardef\leqslant="3\AMSa36
      \mathchardef\geqslant="3\AMSa3E

      \let\leq=\leqslant 

    \fi
  \fi
\fi 
 
\ifnfsstwo
  \newcommand{\rmn}[1] {\mathrm{#1}}

  \DeclareMathAlphabet{\mathbfit}{OT1}{cmr}{bx}{it}
  \SetMathAlphabet\mathbfit{bold}{OT1}{cmr}{bx}{it}
  \DeclareMathAlphabet{\mathbfss}{OT1}{cmss}{bx}{n}
  \SetMathAlphabet\mathbfss{bold}{OT1}{cmss}{bx}{n}
  \ifAMStwofonts
    \ifCUPmtlplainloaded \else
      \DeclareSymbolFont{UPM}{U}{eur}{m}{n}
      \SetSymbolFont{UPM}{bold}{U}{eur}{b}{n}
      \DeclareSymbolFont{AMSa}{U}{msa}{m}{n}
      \DeclareMathSymbol{\upi}{0}{UPM}{"19}
      \DeclareMathSymbol{\umu}{0}{UPM}{"16}
      \DeclareMathSymbol{\upartial}{0}{UPM}{"40}
      \DeclareMathSymbol{\leqslant}{3}{AMSa}{"36}
      \DeclareMathSymbol{\geqslant}{3}{AMSa}{"3E}

      \let\leq=\leqslant 

    \fi
  \fi
\fi 
 
\ifCUPmtlplainloaded \else
  \ifAMStwofonts \else 
    \def\upi{\pi}
    \def\umu{\mu}
    \def\upartial{\partial}
  \fi
\fi

\input psfig.sty
\input epsf.sty

\newcommand\th{\thinspace}
\newcommand\kms{\ifmmode{\rm km\th s^{-1}}\else km\th s$^{-1}$\fi}
\newcommand\ms{\ifmmode{\rm m\th s^{-1}}\else m\th s$^{-1}$\fi}
\newcommand\msun{\ifmmode{M_{\odot}}\else $M_{\odot}$\fi}
\newcommand\rsun{\ifmmode{R_{\odot}}\else $R_{\odot}$\fi}
\newcommand\mjup{\ifmmode{M_{\rm J}}\else $M_{\rm J}$\fi}

\title[Multiple stellar systems -- IV. Gliese 644]
      {Studies of multiple stellar systems -- IV. The triple-lined
spectroscopic system Gliese 644}
\author[T. Mazeh et al.] {Tsevi Mazeh$^1$, David W. Latham$^2$, 
Elad Goldberg$^1$, 
\newauthor Guillermo Torres$^2$,  Robert P. Stefanik$^2$, 
Todd J. Henry$^{2,3}$, 
\newauthor Shay Zucker$^1$, Orly Gnat$^1$ and Eran O. Ofek$^1$
\thanks{Some of the observations reported here were obtained with
the Multiple Mirror Telescope, a joint facility of the
Smithsonian Institution and the University of Arizona.} \\
  $^1$School of Physics and Astronomy, Tel Aviv University, Tel Aviv, Israel\\
  $^2$Harvard-Smithsonian Center for Astrophysics, 60 Garden Street,
  Cambridge, MA, 02138, USA\\
  $^3$Present address: Department of Physics
  and Astronomy, Georgia State University, 1 Park Place Atlanta, 
  GA  30303-3083}

\date{Submitted July 28 2000; Accepted February 15 2001}

\pagerange{\pageref{firstpage}--\pageref{lastpage}}
\pubyear{2000}
\begin{document}

\maketitle

\label{firstpage}

\begin{abstract}
We present a radial-velocity study of the triple-lined system Gliese 644
and derive spectroscopic elements for the inner and outer orbits with
periods of 2.9655 and 627 days.  We also utilize old visual data, as
well as modern speckle and adaptive optics observations, 
to derive a new astrometric
solution for the outer orbit. These two orbits together allow us to
derive masses for each of the three components in the system:
$M_{\rmn{A}} = 0.410 \pm 0.028$ (6.9\%), $M_{\rmn{Ba}} = 0.336 \pm
0.016$ (4.7\%), and $M_{\rmn{Bb}} = 0.304 \pm 0.014$ (4.7\%) \msun.  
We suggest that the relative inclination of the two orbits is very
small.
Our
individual masses and spectroscopic light ratios for the three M stars
in the Gliese 644 system provide three      points for the
mass-luminosity relation near the bottom of the Main Sequence, where the
relation is poorly determined.  These three points agree well with
theoretical models for solar metallicity and an age of 5 Gyr.  Our
radial velocities for Gliese 643 and vB 8, two common-proper-motion
companions of Gliese 644, support the interpretation that all five M
stars are moving together in a physically bound group.  We discuss
possible scenarios for the formation and evolution of this
configuration, such as the formation of all five stars in a sequence of
fragmentation events leading directly to the hierarchical configuration
now observed, versus formation in a small N cluster with subsequent
dynamical evolution into the present hierarchical configuration.
\end{abstract}

\section{INTRODUCTION}

This paper is the fourth in a series on triple-star systems.  The
overall goal of the series is to contribute to our understanding of the
formation and evolution of multiple-star systems.  Paper I (Mazeh,
Krymolowski \& Latham 1993) presented an orbital solution for the
single-lined spectroscopic triple star G38-13.  Paper II (Krymolowski \&
Mazeh 1999) developed an analytical second-order approximation for the
long-term modulation of the orbital elements of triple systems. Paper
III (Jha et al.\ 2000) analyzed the triple-lined system HD 109648,
presenting observational evidence for such modulations.  The present
paper is devoted to the nearby triple-lined system Gliese 644.

Gliese 644 (=Wolf~630=HD~152751=HIP~82817; $\alpha$=16:55:28.76,
$\delta$=$-$08:20:10.8 [J2000], $V=9.02$ mag) is a nearby system of M
dwarfs at a distance of about 6 pc (Gliese 1969).  The study of its
multiplicity began when Kuiper (1934) discovered that Gliese 644 is a
visual binary, later found to have a period of 1.7 years and
semi-major axis of $0\th\farcs218$ (Vo\^ute 1946). Joy (1947) noticed
large radial-velocity variations for Gliese 644, which led him to
suggest that one of the two visual components is itself a
spectroscopic binary with a period of a few days, making Gliese 644
one of the nearest triple systems. 

Weis (1982) used photographic plates to derive a photocentric orbit
and confirmed Fleischer's (1957) suggestion that the fainter component
of the visual binary, B, is more massive than the primary.  He
concluded that B, which is about 0.1 mag fainter than A in the visual,
is the short-period spectroscopic binary. Weis derived masses of
$M_{\rmn{A}} = 0.28$ and $M_{\rmn{B}} = 0.56\ \msun$, suggesting
that the system is composed of three similar M dwarfs.  These masses
are consistent with the dM2.5 spectral type assigned by Henry,
Kirkpatrick \& Simons (1994) to the blended image of Gliese~644.

In order to learn more about Gliese 644, we started fifteen years ago
to monitor the object spectroscopically. This was done within a
radial-velocity study of a small sample of nearby M dwarfs carried out
with the facilities at the Harvard-Smithsonian Center for Astrophysics
(CfA).  From the beginning of the project we could see two and
occasionally even three peaks in some of the one-dimensional
cross-correlation functions, which were obtained by correlating the spectra of
Gliese 644 against our standard observed M-star template.  Given the
visual orbit of the outer binary, the triple-lined nature of Gliese
644 makes this system special, as it enables us to derive individual
masses for each of the three components if spectroscopic solutions can
be derived for both the inner and the outer orbits.  This can add
important information about the mass-luminosity relation in the solar
vicinity (cf. Andersen 1991, S\"oderhjelm 1999) near the bottom of the
Main Sequence (e.g., Henry et al.\ 1999). 

The derivation of radial velocities for all three components in the
Gliese 644 system presented an unusually difficult challenge, because
the inner and outer orbits both have small radial-velocity amplitudes,
resulting from the nearly face-on inclination of the outer orbit, $i
\sim 165\degr$, from the relatively long period of the outer orbit, $P
\sim 1.7$ yr, and from the relatively small masses of the three stars,
$M \sim 0.3$ \msun. The lines from all three stars are rarely resolved
in our spectra; in most cases all three are blended together. To solve
the problem of extracting velocities for all three stars from blended
spectra, we had to wait for the development of a more powerful algorithm
than the one-dimensional cross-correlation techniques that we had been
using. Such an algorithm is TODCOR (Zucker \& Mazeh 1994), which was
developed originally as a two-dimensional correlation technique for
extracting velocities for both stars in double-lined spectroscopic
binaries, even when the two sets of lines were not resolved.  The
extension of TODCOR to three dimensions for the analysis of triple-lined
systems (Zucker, Torres \& Mazeh 1995) provided us with the tool that
we needed to analyze the spectra of Gliese 644.

Even with TODCOR, the extraction of radial velocities for all three
components of Gliese 644 proved to be difficult.  To address this
difficulty, we developed a new approach for solving spectroscopic
orbits for systems with composite spectra.  The new approach searches
a model that includes the observed spectra {\it together} with the
orbital parameters of the system. The final model is the one that
gives the best match between the observed spectra and the
corresponding set of composite spectra predicted by the model.

The observations were analyzed with this approach by two different
teams, one at Tel Aviv and the other at the CfA, with some differences
in the details at various stages of the analysis.  The two analyses
led to similar results, and most of the orbital elements agreed within
the internal uncertainty estimates.  However, the velocity amplitudes for
the outer orbit disagreed by 3.2 and 3.5 times the internal error
estimates, suggesting the presence of significant {\it systematic}
difference between the two analyses. Therefore, we present the
results from both analyses and document the procedures used in detail.
For the final orbital parameters we adopt simply the averages of the
values derived by the two analyses.

Visual observations of the wide pair Gliese 644AB extend back more than
50 years.  Because some recent unpublished speckle observations were
available to us, we decided to reanalyze the astrometric orbit of the
outer binary. Comparison between the results of the spectroscopic and
astrometric analyses indicates some possible systematic scale
differences between the two sets of data. This impression was supported
by the observations and analysis of Gliese 644 by S\'egransan et
al. (2001), a work published about three months after we submitted our
paper, while our paper was still being reviewed. Their astrometric
orbit, which was based solely on speckle and newly obtained adaptive
optics (AO) measurements and did not include any visual observations,
yielded a somewhat larger semi-major axis. We find a similar trend when
we use only the modern astrometric observations to estimate the
semi-major axis. Because of the importance of the scale of the
astrometric orbit for the mass determinations, and because the visual
observations are more likely to be afflicted by systematic errors, we
decided to rely only on the modern data to set the scale of the orbit.

The details of the visual analysis are reported in section 2.
A new approach for solving spectroscopic orbits of systems with
composite spectra is described in Section 3, together with the details
of the Tel Aviv and CfA orbital solutions. In Section 4 the solutions
for the spectroscopic and astrometric orbits are combined to solve for
the individual masses of the three M stars in the Gliese 644 system.
Section 5 discusses in detail the visual and IR photometry we have in
hand for the A and B components of the system, and Section 6 considers
the impact of our results on the mass-luminosity relation near the
bottom of the Main Sequence.  Section 7 discusses the relationship of
Gliese 644 to Gliese 643 and vB 8, two nearby common-proper-motion
companions.  Our radial-velocity measurements support the interpretation
that all five stars are moving together in a physically bound system.
In the final section we summarize our results and discuss the
implications for two scenarios describing the formation and evolution of
binary and multiple systems.

\section{THE ASTROMETRIC ORBIT FOR THE OUTER BINARY}

\subsection{The visual and speckle data}

For nearly 50 years Gliese 644AB was the shortest-period (1.7 yr)
binary resolved by visual means, with an angular separation of
$\sim$0\th\farcs2. Only when interferometric and speckle techniques
became available was it possible to resolve systems with even smaller
separations.  Consequently, Gliese 644AB attracted a great deal of
attention, and a large number of visual observations have accumulated
since its discovery. 

\begin{figure}
\vspace{-0.5in}
\def\epsfsize#1#2{0.45#1}
\epsfbox{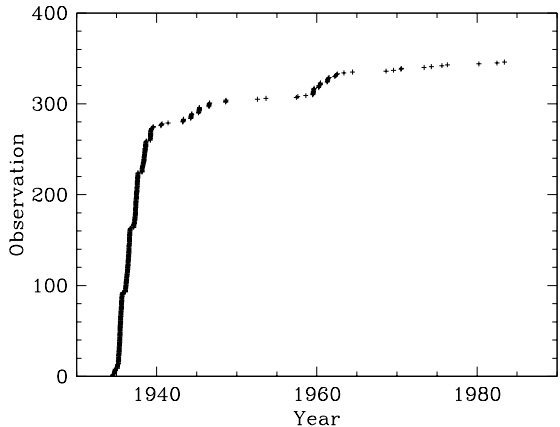}
\vspace{-0.5in}
\caption{Cumulative distribution of the visual observations.}
\label{astrometry.cumulative}
\end{figure}

The history of the astrometric observations of Gliese 644AB is far
from uniform. During the first five seasons after its discovery more
than 270 visual measurements were made, approximately 200 of them by
the same observer, J.\ G.\ Vo\^ute, at the
Bosscha Observatory in Lembang, Java. Vo\^ute was also the first to
publish an orbit for the system (Vo\^ute 1946). In the following 45
years some 80 additional visual observations were collected by a number of
observers. The cumulative distribution of all the visual observations
is shown in Figure \ref {astrometry.cumulative}, where the irregular
pattern is obvious. 

In recent years speckle observations of Gliese 644AB have also been
made at both visual and infrared wavelengths, providing additional
measurements of both the separation and the relative brightness of the
two wide components.  Three of the speckle measurements were obtained
by one of us (Henry 1991) in the infrared ($J$, $H$, $K$), using the
Steward Observatory 2.3-m telescope on Kitt Peak, and have not
appeared in the literature. Two are one-dimensional scanning
observations in the North-South direction, and one is a
two-dimensional observation.  Descriptions of the instrumentation and
observing techniques can be found in McCarthy (1986) for the
one-dimensional speckle observations, and McCarthy et al.\ (1991) for
the two-dimensional speckle observation. Finally, S\'egransan et al.\
(2001) have very recently reported five adaptive optics measurements
that are of very high precision. 

A listing of all the astrometric observations, with the exception of
the three speckle measurements mentioned above and the AO
measurements, was provided to us by the late Charles E.\ Worley (U.S.\
Naval Observatory) and was extracted from the Washington Double Star
Catalog (Worley \& Douglass 1996).  The three unpublished speckle
observations are listed in Table \ref{astrometry.tab}.  The type of
measurement is indicated by ``2ds'' (two-dimensional speckle) or
``1ds'' (one-dimensional speckle). 

\begin{table}
\caption{Unpublished speckle observations for Gliese 644AB.}
\begin{tabular}{lcccc}
\hline
\phantom{I(}Date & Orbital & $\theta$ & $\rho$ & \\
\phantom{I}(Year) & Phase & (deg) & (arcsec) & Type \\
\hline

  1984.3570 &  0.797  &         &           0.13  &  1ds \\
  1989.2220 &  0.631  &         &           0.17  &  1ds \\
  1990.3530 &  0.289  &   301   &\phantom{0}0.198 &  2ds \\

\hline
\end{tabular}
\label{astrometry.tab}
\end{table}

With this wealth of information covering nearly 40 orbital cycles
(1934-2000) it should in principle be straightforward to obtain a
high-quality solution for the visual orbit. Indeed, a number of
solutions have been published since Vo\^ute's first determination
(e.g., Starikova 1980, Heintz 1984), although the original orbit is
the one listed as ``definitive'' in the Fourth Catalog of Orbits of
Visual Binary Stars (Worley \& Heintz 1984). The combination of the old
and the new data can in principle improve the orbital solution.
This prompted us to take the present study of Gliese 644 as
an opportunity to rediscuss the visual orbit.

\subsection{The orbital solution}

Using all the available astrometric observations
together raises two problematic issues. The first has to do with the
variety of observers, observing conditions, and techniques used to
collect the astrometric data since 1934, which could give rise to {\it
systematic} errors.  The second and more fundamental issue is
associated with the triple nature of Gliese 644, which introduces
tidal interactions between the outer and inner orbits. This could
produce slow changes in the elements of the outer orbit (as well as
the inner orbit) on long timescales (e.g.  Mazeh \& Shaham 1976,
1977). The timescale for a complete cycle of such a modulation is of
the order of 700 years (Mazeh \& Shaham 1979), while the actual
timescale could be shorter.  Thus we cannot rule out {\it a priori}
some small changes in the elements of the wide orbit during the 66
years of observational coverage. 

In order to address the first of these issues, we begin by pointing
out a potentially serious shortcoming of the observations obtained by
the visual technique. These data are dominated in number by the
measurements made by Vo\^ute, who observed with a 60-cm refractor. At
visual wavelengths the diffraction limit of such an instrument is
approximately 0\farcs23, which is essentially the same as the angular
separation displayed by Gliese 644AB throughout its entire orbit. In
his original publications Vo\^ute reports that the measurements are
frequently only ``estimates" made from elongated or notched images.
Under these conditions angular separation measurements by the visual
technique must be taken with extreme caution, since they have often
been shown to be biased (see, e.g., Douglass \& Worley 1992) and could
affect the semimajor axis of the orbit. Nevertheless, the visual
observations are potentially valuable for the extended time coverage
they provide.  Consequently, we started by considering a solution that
incorporates only the visual measurements
and none of the modern speckle or AO observations,
which could have a different scale. Evidence of just this effect is
discussed below.

To compute the orbital solutions described in this section we used
standard non-linear least-squares techniques based on the
Levenberg-Marquardt method (Press et al.\ 1992). All position angles
were precessed to the epoch 2000.  The solutions were obtained
iteratively, solving for an orbit and adopting the root mean square
residual as the uncertainty for the next iteration, until convergence.
The errors determined through this procedure are 0\th\farcs012 and
0\th\farcs013/$\rho$, which are surprisingly good for the visual
technique. Eight observations gave unusually large residuals in these
preliminary solutions ($\gg3\sigma$), and were rejected. Observations in
which only the angular separation was discrepant were not entirely
rejected, because the position angles can still be accurate (they are
measured essentially independently from the separations in both the
visual and the speckle techniques).  These observations can contain
valuable information on the period of the system, and should be included
in the solution.

\begin{figure}
\vspace{-0.7in}
\def\epsfsize#1#2{0.45#1}
\epsfbox{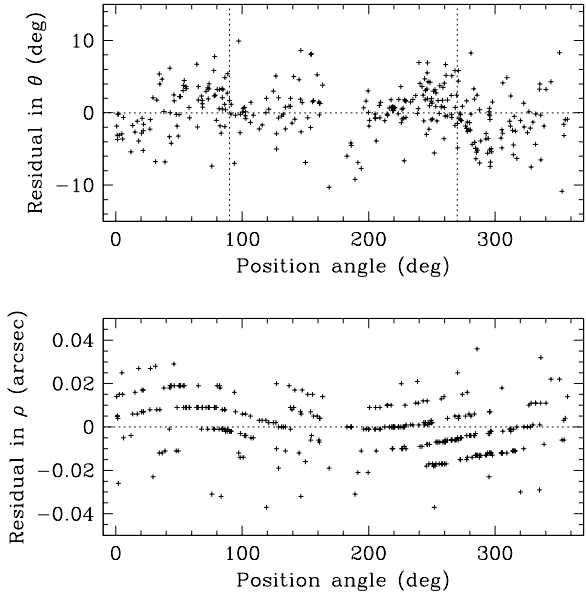}
\vspace{-0.7in}
\caption{Residuals in $\theta$ and $\rho$ from an astrometric orbit that
uses only visual data, shown as a function of position angle.}
\label{astrometry.residuals}
\end{figure}

The residuals from a solution using only the visual data are shown in
Figure \ref{astrometry.residuals}. Unusual trends are seen in the
residuals of both $\theta$ and $\rho$, plotted here as a function of
position angle in the orbit.  The bottom panel shows an obvious
pattern of parallel curves separated by exactly 0\th\farcs01, which is
the precision to which visual observers typically read off the
separations from their filar micrometers. A less obvious effect is
seen in the $\theta$ residuals displayed in the top panel, which seem
to change sign on either side of 90\degr\ and 270\degr\ (shown as
vertical dotted lines). In addition, when plotted as a function of
time (not shown here), the residuals in $\theta$ are systematically
negative by about 5\degr\ during 1935, which was the first of
Vo\^ute's very intensive observing seasons for Gliese 644. 

To further investigate the latter effect we performed a sequence of
additional orbital solutions in which we removed the older
measurements one by one and recomputed the orbital elements in each
case. These fits are therefore not independent, and perhaps a better
way would be to compute solutions in a window corresponding to a fixed
interval of time or a fixed number of observations, sliding it along
the time axis. Unfortunately the peculiar time history for the
astrometry of Gliese 644, with more than 75\% of the data obtained in
the initial 10\% of the time coverage, does not allow this (see Figure
\ref{astrometry.cumulative}).  As a result of this exercise, we
detected sharp changes in the orbital elements as soon as the
observations in 1935 are excluded. Although tidal effects as discussed
earlier can produce long-term changes in the orbital elements, the
discontinuities observed are much to abrupt to be real. An alternate
explanation would be a systematic error in the observations for 1935,
the vast majority of which were made by Vo\^ute. From the discussion
above it appears likely that his position angles include a systematic
error. This is somewhat unexpected for such a careful and experienced
observer as Vo\^ute, particularly since his later observations do not
show this effect.  There are no clues on the source of the systematic
error, despite the fact that he recorded his observations in great
detail, and we can only speculate as to the cause. We have therefore
chosen to exclude Vo\^ute's first observing season altogether, after
which the discontinuities in the orbital elements mentioned above
disappear completely. 

The remaining visual observations, which cover nearly 50 years, were
used to examine the possibility of changes in the orbital elements due
to tidal effects. For this we divided the measurements into two
independent groups, although once again as seen in Figure
\ref{astrometry.cumulative} the time sampling is not very favorable
for this sort of test. As a compromise between the number of orbital
cycles spanned and the number of observations in each group, we chose
1955 as the dividing point. Independent solutions using the pre-1955
and post-1955 data are given in Table \ref{astrometry.elements}. 


\begin{table*} 
\caption{Astrometric orbital solutions for Gliese 644AB,
excluding the 1935 observations by Vo\^ute.}
\begin{tabular}{lccccc}
\hline
Element & Pre-1955 & Post-1955 & All visual & Modern & Adopted\\
        & (visual only) & (visual only) &  & (no visual)\\
\hline
$P$ (yr)                   & $1.71576 \pm 0.00058$ & $1.71634 \pm 0.00088$ & $1.71687 \pm 0.00018$ & $1.71671 \pm 0.00091$  & $1.717267 \pm 0.000039$\\
$P$ (d)                    &  $626.67 \pm 0.21$    &  $626.89 \pm 0.32$    & $627.087 \pm 0.066$   & $627.03 \pm 0.33$      & $627.232 \pm 0.014$   \\
$a (\arcsec)$              &  $0.2169 \pm 0.0013$  &  $0.2143 \pm 0.0035$  &  $0.2166 \pm 0.0013$  &  $0.2238 \pm 0.0023$   & $0.2256 \pm 0.0011$  \\
$e$                        &  $0.0337 \pm 0.0026$  &  $0.0666 \pm 0.0082$  &  $0.0384 \pm 0.0026$  &  $0.0448 \pm 0.0066$   &  $0.0433 \pm 0.0018$ \\
$i (\degr)$                &   $165.0 \pm 1.7$     &   $163.2 \pm 5.1$     & $164.9 \pm 1.6$       &   $166.9 \pm 6.1$      &  $163.1 \pm 1.6$   \\
$\omega (\degr)$           &   $124.4 \pm 7.7$     &      $83 \pm 17$      &  $118.9 \pm 7.2$      &    $98   \pm 40 $      &  $115.6 \pm 5.1$  \\
$\Omega_{2000} (\degr)$    &   $169.7 \pm 6.4$     &     $140 \pm 17$      &  $167.2 \pm 6.3$      &   $150   \pm 26 $      &  $163.2 \pm 3.1$  \\
$T$                        &$1988.111 \pm 0.024$   &$1988.083 \pm 0.036$   &$1988.129 \pm 0.017$   &$1988.122 \pm 0.067$    & $1988.143 \pm 0.011$  \\
$a^3/P^2 ({\rmn{arcsec}}^3 {\rmn{yr}}^{-2})$&$0.003468\pm 0.000058$& $0.00334 \pm 0.00021$ &  $0.003446 \pm 0.000063$&$0.00380 \pm 0.00020$  &  $0.003893 \pm 0.000058$\\
\\
$N_{\theta}$               &          231          &           39          & 270   &   14           &         284  \\
$N_{\rho}$                 &          219          &           37          & 258   &   13           &13 \\
$N_{\rmn 1-D}$             &            0          &            0          &   0   &    1           &1 \\
Span (yr)                  &         19.1          &         26.0          & 48.9  & 16.0           &66.0 \\
\hline
\end{tabular}
\label{astrometry.elements}
\end{table*}

The orbital elements derived from the two data sets are rather
similar, with the exception of the eccentricity. The change in $e$
amounts to 3.8 times the combined errors, an apparently significant
effect.  Interestingly enough, this is one of the parameters expected
to vary due to the tidal interaction between the outer and inner
orbits, and for some triples it displays the most prominent modulation
(Mazeh \& Shaham 1979; Paper II). Although very suggestive, we
hesitate to place much confidence in this result at the present time
in view of the systematic trends in the visual data illustrated in
Figure \ref{astrometry.residuals}, which could alter the shape of the
orbit in subtle ways. Nevertheless, continued observations with modern
techniques may well confirm this in the future. A combined solution
using all the visual data together is shown in the third column of the
table. 

Because of the potential for systematic errors in the angular
separations of the visual data, a completely independent solution was
carried out using only the modern speckle and AO measurements.  The
one-dimensional speckle observations also contain useful information on
the angular separation, and can be included in the fit as well.  We
assigned errors for the (two-dimensional) speckle measurements of
0\th\farcs01 in angular separation ($\rho$) and 0\th\farcs01/$\rho$ in
position angle ($\theta$). For the one-dimensional speckle measurements
we adopted an error of 10~percent.  We note, however, that the
uncertainties reported by S\'egransan et al.\ (2001) for their five AO
measurements are substantially smaller than those of the speckle
measurements, with the result that they carry a very large weight in
this new fit. Two of those observations, in particular, have quoted
errors in the angular separation of only 0\farcs0005. If allowed to
carry this enormous weight (400 times that of the speckle technique),
these AO measurements would completely dominate the solution, a risky
situation when little is known about the {\it systematic\/} errors they
may be affected by. We have preferred to be conservative, and have
therefore chosen to set the errors in angular separation of these two AO
observations to the average of the other three, which is 0\farcs003.

 The new fit based
exclusively on modern observations is shown in the fourth column of
Table \ref{astrometry.elements}. Compared to the solution that uses
all the visual data, uncertainties in all the elements are larger due
to the smaller number of observations and the shorter time coverage.
Still, the elements of the two fits are consistent within
1$\sigma$, with the exception of the semimajor axis (2.7$\sigma$),
which is larger in the modern fit than indicated by the visual
solution. 

Similar indications come from a semi-independent solution published
for the astrometric orbit based on data from the Hipparcos mission.
Gliese 644AB was one of the targets observed by the satellite
(HIP~82917), and although the orbital elements are not actually listed
in the Hipparcos Catalogue (ESA 1997), a re-analysis of the
intermediate transit data was performed by S\"oderhjelm (1999) as part
of a study of 205 binary systems. Ground-based observations were
incorporated to some extent in S\"oderhjelm's analysis, so that his
solution for Gliese 644AB is not completely independent of ours.
Nevertheless, the semimajor axis he obtained, $a = 0\farcs23$, is
larger than suggested by the visual data, possibly due to his reliance
on the very precise satellite measurements to set the scale. 

As mentioned earlier, many of the angular separations measured
visually are merely estimates that were made at or under the
resolution limit of the telescope, and as such they are particularly
susceptible to personal (i.e., subjective) errors. It seems safer,
therefore, to rely on the modern speckle and adaptive optics
measurements to set the scale, since these are considerably more
objective in nature. On the other hand, the position angles of the
visual measurements may still be useful since they are typically more
accurate than the angular separations (see, e.g., Pannunzio et al.\
1986). The optimal solution for Gliese 644AB, therefore, is one that
uses the modern data (both $\theta$ and $\rho$) mainly for scale, and
the visual position angles that provide the time coverage to
strengthen the orbital period. Subtle effects such as those
illustrated in the top panel of Figure \ref{astrometry.residuals} will
tend to cancel out over many cycles and will not otherwise affect the
solution significantly because of the much smaller weight of the
visual data. The result of this combined fit is given in the last
column of Table \ref{astrometry.elements}, and these are the orbital
elements we adopt for the remainder of this paper. 

\begin{figure}
\def\epsfsize#1#2{0.45#1}
\epsfbox{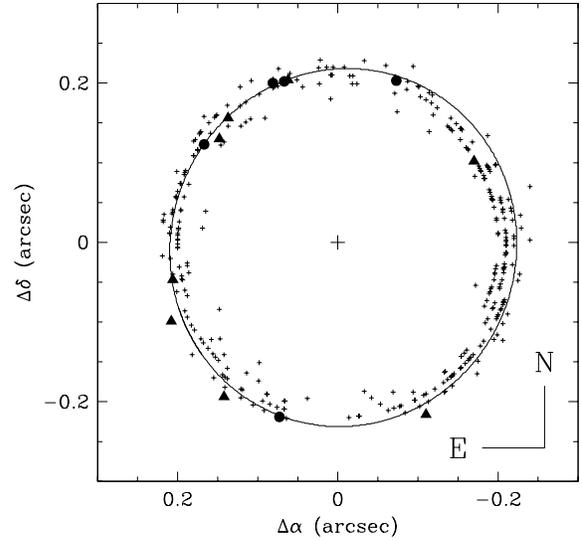}
\caption{The astrometric orbit and the observations on the plane of
the sky.  The scale is set solely by speckle and AO measurements (see
text). The plus signs represent the visual observations, and the
triangles and circles are for the speckle and AO measurements,
respectively.  Motion is clockwise.}
\label{astrometry.orbit}
\end{figure}

A graphical representation of the orbit and the observations on the
plane of the sky are shown in Figure~\ref{astrometry.orbit}. As
expected, the visual observations fall mostly {\it inside} the orbit,
since the angular separations are typically underestimated, as
discussed above. 

Our new orbital solution is in good agreement with the old astrometric
solutions, except for the scale. The plane of the orbit is highly
inclined (only $15\degr$ from face on), the motion is retrograde, and
the orbit is nearly circular, with a very small but significant
eccentricity.  When only an astrometric orbit is available, the values
of $\omega$, the longitude of the periastron, and $\Omega$, the
position angle of the ascending node, are both ambiguous by
$180\degr$.  The values quoted in Table \ref{astrometry.elements}
eliminate this ambiguity by taking into account the spectroscopic
orbit presented below.  In particular, the relative motion of the
secondary is away from the Sun at the ascending node. 

The inclination angle we derive, $i = 163\fdg1 \pm 1\fdg6$, is
slightly larger than the values by S\'egransan et al.\ (2001) ($i =
160\fdg3 \pm 0\fdg1$) and S\"oderhjelm (1999) ($i = 161\degr$),
although in the latter case no errors are given and it is difficult to
evaluate the results.  On the other hand, the eccentricity found by
S\"oderhjelm (1999), which presumably corresponds on average to a
fairly recent epoch because of the use of the Hipparcos measurements,
is $e = 0.06$, which is very close to the value we obtain from the
post-1955 visual data. This would seem to support the difference we
noted earlier, but until stronger evidence is found all we can say at
the moment is that the ground-based and Hipparcos data cannot rule out
a possible slight increase in the eccentricity of the outer orbit of
Gliese 644 on a long timescale.

\section{THE SPECTROSCOPIC ORBITS FOR THE INNER AND OUTER BINARIES}

We have monitored the spectrum of Gliese 644 with the CfA Digital
Speedometers (Latham 1985, 1992), using three nearly identical
instruments.  Most of the spectra were obtained with the 1.5-m
Tillinghast Reflector at the F.\ L.\ Whipple Observatory atop Mt.
Hopkins, Arizona (113 exposures) and the 1.5-m Wyeth Reflector located
at the Oak Ridge Observatory in the town of Harvard, Massachusetts (77
exposures).  A few of the spectra were obtained with the MMT, also
located at the Whipple Observatory (10 exposures).  Photon-counting
intensified Reticon detector systems were used to record a single
echelle order centered near 5187 \AA, with a spectral coverage of 45
\AA\ and resolution of 8.3 \kms.  The signal-to-noise ratio ranged from
7 to 40 per resolution element in the continuum, with 20 being a
typical value.  The first 194 exposures were obtained over a 2645-day
span between 1984 and 1991, while the final 6 exposures were obtained
10 years later in 1997. 

To analyze the data we developed a new approach in which a candidate
model for the system and its two spectroscopic orbits is judged by how
well it reproduces the set of observed spectra. For the sake of
simplicity we choose to discuss our approach only for the case of a
double-lined binary.  Extending our approach to the triple-lined case is
straightforward.

\subsection{The New Approach}

\subsubsection{The old procedure}

In previous studies of double-lined systems (e.g., Mazeh et al.\ 1995,
Goldberg et al.\ 2000) we have derived orbital solutions in two
separate steps. In the first step we extracted individual radial
velocities for the two stars from each of the observed spectra.  Then
in the second step we derived the orbital parameters, based on the
velocities obtained in the first step. The velocity extraction was
accomplished using TODCOR, a two-dimensional correlation technique
(Zucker \& Mazeh 1994) that assumes the observed spectrum is a
combination of two spectra, with shifts caused by the radial
velocities of the two components. The algorithm calculates the
correlation of the observed spectrum against a combination of two
templates with different shifts. The result is a two-dimensional
correlation function, whose maximum is expected at the shifts
corresponding to the actual velocities of the two components.  To find
the orbital solutions of the binary systems we used ORB (Paper I) --- a
code that searches the orbital parameter space for a minimum of the
residuals of the stellar radial velocities. 

To derive the radial velocities of the two stars TODCOR requires two
template spectra, one for the primary and the other for the secondary.
To match the actual components of each binary we chose templates from
a library of synthetic or observed spectra.  To optimize the match
between the templates and the observed composite spectra, we ran
TODCOR for a variety of different templates from the library. To each
pair of templates we assigned a measure of its fit to the spectra of
the binary -- the peak correlation obtained by TODCOR for that pair of
templates, averaged over all the observed spectra.  We then chose the
pair of templates that gave the highest average value for the peak. 

To extract the radial velocities for the two stars, the brightness
ratio of the two templates, $\alpha$, is also needed.  For each
observed spectrum TODCOR can derive the radial velocity of the two
components either by accepting a predetermined brightness ratio or by
finding the best ratio to fit that spectrum at the velocities found.
Normally we assumed the binary system had the same brightness ratio
for all the observed spectra, and therefore adopted a single value of
$\alpha$ for all the spectra. To choose the best brightness ratio we
either averaged the derived $\alpha$ values over the observed
spectra, or ran TODCOR for a grid of predetermined $\alpha$ values,
searching for the value that gave the highest average value for the
peak. 

Choosing the best templates and best value for $\alpha$ comprised the
major part of the first step of the old procedure.  Only after the
templates and the brightness ratio had been established did we proceed
to derive the radial velocities, which were then used by ORB to find
the orbital solution. 

\subsubsection{A modified approach}

The separation of the reduction procedure into the two steps described
above is somewhat artificial and is not necessarily the optimum way to
find the best model for the system. We describe here a new approach,
based on the philosophy that the model as a whole, namely the orbital
elements {\it together} with the choice of templates and brightness
ratio, should be confronted directly with the observed spectra. In
short, we propose that for any suggested model of the system one should
construct a set of predicted spectra and compare them with the observed
spectra.
 
This confrontation can be done in a straightforward way. Consider a
model consisting of a given set of values for the orbital parameters
{\it together} with a choice of templates and brightness ratio. First,
derive two stellar velocities predicted by the orbital parameters for
the timing of each of the observations. Second, construct a {\it
predicted spectrum} for each observation by shifting the templates
with the predicted velocities and combining the two shifted templates
with the adopted brightness ratio. To compare the predicted spectrum
with the observed one, calculate the correlation between the two
spectra.  Finally, the correlations of the different observed spectra
should be combined into a global score, which measures the match of
all the spectra to the model.

Note that in this new approach the correlation calculated for an
individual observed spectrum is the correlation at the velocities
predicted by the orbital solution. This is not necessarily the highest
possible correlation for that observed spectrum.  There might be a
different pair of velocity shifts for the two templates that yields a
higher correlation. However, in the new approach what counts is the
correlation corresponding to the two velocities predicted by the
specific orbital solution, calculated for the time of the observation. 

The best model is the one that yields the highest correlation
score.  Therefore, the algorithm we propose performs a search for the
best model in the complete parameter space, which includes the orbital
parameters and the possible templates and brightness ratios. To perform
an efficient search we
suggest an iterative procedure. The solution found
in the old approach serves as our first guess. This gives us a choice
of templates and brightness ratio, together with an orbital solution.
We then iterate, so that every stage of the iteration consists of
three steps:

\begin{itemize}

\item A search for the best templates and brightness ratio, {\it
given} the velocities predicted by the previous orbital solution. 

\item A new TODCOR derivation of stellar velocities, {\it given} the
new templates and brightness ratio. 

\item The determination of a new orbital solution with ORB, by
minimizing the $\chi^2$ of the residuals with regard to the new
velocities. 

\end{itemize}

We require that each step yields a higher correlation
score, and that the iterations converge. 

The extension of the new approach to triple-lined systems is
straightforward. We need to search for three templates and two
brightness ratios, $\alpha$ and $\beta$, one corresponding to the
ratio between the secondary and the primary brightness, the other to
the ratio between the tertiary and the primary. 

\subsection{Analysis}

The lines of the three stars of Gliese~644 are blended in most of our
observed spectra. It is therefore especially important to have good
template spectra, optimized to match each of the three components.  In
particular, it is important that the template spectra have the correct
rotational broadening when using TODCOR to determine radial velocities
from blended spectra.  If there is not enough rotational broadening in
the templates, TODCOR is forced to pick orbital velocities that are too
far apart, in order to match the observed line broadening.  Conversely,
TODCOR will pick orbital velocities that are too close if the templates
have too much rotational broadening.

For most of our radial-velocity projects we use template spectra drawn
from an extensive library of synthetic spectra calculated by Jon Morse
for a large grid of Kurucz model atmospheres (cf. Nordstr\"om et al.\
1994, Morse \& Kurucz in preparation).  Synthetic templates have the
advantage that they can be calculated for dense grids in effective
temperature, surface gravity, metallicity, and rotational broadening.
If observed spectra are used as templates, it is hard to find real stars
that fill densely such grids.  Unfortunately, the Kurucz models start to
become increasingly unrealistic for effective temperatures below about
4000 K.  For M dwarfs we have found that we get better correlations if
we use templates drawn from a small library of observed spectra with
spectral types in the range K7 to M5.5.  The observed templates
available for our analysis of Gliese 644 are summarized in Table
\ref{templates.tab}, where we list the stellar spectral types of Henry
et al.\ (1994).

\begin{table} \caption{The stars observed for templates.}
\begin{tabular}{cccc}
\hline
Gliese & $\alpha$~~(J2000)~~$\delta$         & $V$   & Spectral \\
       &                                     & (mag) & Type     \\
\hline
380    & 10:11:22.14~~+49:27:15.3            &  6.61 &   K7  \\
809    & 20:53:19.79~~+62:09:15.8            &  8.54 &  M0.0 \\
846    & 22:02:10.27~~+01:24:00.8            &  9.16 &  M0.5 \\
908    & 23:49:12.53~~+02:24:04.4            &  8.98 &  M1.0 \\
15A    & 00:18:22.89~~+44:01:22.6            &  8.07 &  M1.5 \\
49     & 01:02:38.87~~+62:20:42.2            &  9.56 &  M2.0 \\
745B   & 19:07:13.\phantom{22}~~+20:52:36.\phantom{2} & 10.75 &  M2.0 \\
48     & 01:02:32.23~~+71:40:47.3            &  9.96 &  M2.5 \\
725A   & 18:42:46.66~~+59:37:50.0            &  8.91 &  M3.0 \\
725B   & 18:42:46.90~~+59:37:36.6            &  9.69 &  M3.5 \\
273    & 07:27:24.50~~+05:13:32.8            &  9.89 &  M3.5 \\
699    & 17:57:48.50~~+04:41:36.2            &  9.54 &  M4.0 \\
51     & 01:03:12.\phantom{22}~~+62:21:54.\phantom{2} & 13.66 &  M5.0 \\
905    & 23:41:54.0\phantom{2}~~+44:09:32.~            & 12.28 &  M5.5 \\
\hline
\end{tabular}
\label{templates.tab}
\end{table}

Throughout our iterations two sets of templates performed significantly
better than any others we tried: Gliese 725A for the primary and Gliese
725B for the secondary and tertiary, and Gliese 725A for the primary and
Gliese 273 for the secondary and tertiary.  These sets gave essentially
the same peak correlation averages.  We decided to adopt the Gliese 725
templates, because we hoped that this would
minimize systematic errors due to differences in velocity zero points,
metallicity, and rotational velocity.

To find the brightness ratio and the orbital parameters
we have developed two independent procedures, one at Tel Aviv
and the other at the CfA.  Completely independent codes were used
throughout, both for the implementation of TODCOR and for the orbital
solutions. We describe the two procedures in the next two subsections.

\subsubsection{The Tel Aviv Analysis}

To find the brightness ratios of Gliese~644 we ran in each iteration a
grid of $\alpha$ and $\beta$, using all 200 velocities derived in the
previous iteration. To average the different correlations calculated for
the different observed spectra we used a ``generalized correlation
score'' derived by Zucker (in preparation) by applying a
maximum-likelihood approach.  This score can be thought of as a way of
weighting the individual correlations by using the estimated signal-to-noise
ratio of each spectrum.  We finally converged to a value of $\alpha =
0.56 \pm 0.06$ and $\beta = 0.36 \pm 0.04$.

Finding the orbital solutions of a triple-lined system for a given set
of velocities is somewhat more complicated than in the case of a
double-lined binary. This is so because we have to solve for two orbits,
one for the close pair and the other for the center of mass of the close
pair together with the distant third star.  To solve simultaneously for
the two orbits we use the code ORB.20, a slightly modified version of
ORB.18 (Paper I). The solution is reached in a few iterative
steps. First, ORB solves for the orbital elements of the primary, whose
sole motion is within the wide orbit. Second, ORB solves for the orbital
elements of the primary and the secondary as a double-lined system in
the wide orbit, ignoring at this stage the short-period motion of the
secondary. Third, it solves for the elements of the primary and the
secondary as a triple system, taking into account for the first time the
secondary short-period motion. At this stage the wide orbit is
considered as a double-lined orbit, while the short-period orbit is
considered as a single-lined one. Then, in the last stage of the
solution, the tertiary data are taken into account, and the short-period
motion is considered as a double-lined orbit. Each step of the iteration
is used as the starting point for the next one.  Finally, we compute
differential corrections for all the elements of the two orbits together.

It has been our experience that the period is the critical parameter
of the orbital solution, so ORB first searches the orbital parameter
space with a dense grid of fixed values of the orbital period,
covering a range of periods given by the user. For every value of the
period ORB searches for the parameters that minimize $\chi^2$.  This
procedure ensures that we do not miss any local minimum in the
parameter space of the orbital elements. Only when the best period has
been found, does ORB execute an iterative algorithm to find the best
set of parameters to fit the data. 

The velocities derived with TODCOR can encounter two problems, both
resulting from the fact that all observed spectra include some noise. The
problems have to do with the fact that TODCOR derives the three
velocities from the location of the highest peak of the correlation in
the three-dimensional velocity space. One problem occurs when a
spurious peak of the three-dimensional correlation gets randomly
enhanced by the noise and becomes higher than the peak that corresponds
to the actual three velocities.  In such a case TODCOR can chose a
{\it completely} wrong peak. Another problem occurs when TODCOR
switches between the velocities of the three stars. This can happen
when the templates are similar and the spectra have poor
signal-to-noise ratios. Both problems can generate outlier velocities
with large residuals. 

To minimize the problem of identifying a completely wrong peak we let
TODCOR search for a peak in a limited region of the three dimensional
space, centered on the point which corresponds to the three velocities
predicted by the orbital solution from the previous iteration.  Usually
a range of $\pm 35$ \kms\ in each dimension is searched.  To identify
velocities switched by TODCOR we search for exposures where the three
derived velocities yield a high $\chi^2$ when compared to the velocities
predicted by the orbital solution. We then switch the velocities if and
only if the switched velocities give a significantly smaller $\chi^2$ for
that exposure.  After switching velocities we then iterate the entire
orbital solution again and require that the new solution give a lower
$\chi^2$.

The
individual radial velocities and (O-C) residuals from the orbital
solutions are reported in Table \ref{radial.velocities.tab}.

\begin{table*}
\caption{Radial velocities and residuals (km s$^{-1}$) for Gliese 644
(first 20 lines)}
\begin{tabular}{lrcrcrcrcrcrc}
\hline
\phantom{IIII}HJD & \multicolumn{4}{c}{\underline{Gliese 644A}} &
\multicolumn{4}{c}{\underline{Gliese 644Ba}} &
\multicolumn{4}{c}{\underline{Gliese 644Bb}} \\ (2,400,000$+$) &
\multicolumn{2}{c}{CfA} & \multicolumn{2}{c}{Tel Aviv} &
\multicolumn{2}{c}{CfA} & \multicolumn{2}{c}{Tel Aviv} &
\multicolumn{2}{c}{CfA} & \multicolumn{2}{c}{Tel Aviv} \\
& $V_{\rm{r}}$\phantom{I} & (O $-$ C) & $V_{\rm{r}}$\phantom{I} & (O
$-$ C) & $V_{\rm{r}}$\phantom{I} & (O $-$ C) & $V_{\rm{r}}$\phantom{I}
& (O $-$ C) & $V_{\rm{r}}$\phantom{I} & (O $-$ C) &
$V_{\rm{r}}$\phantom{I} & (O $-$ C) \\
\hline
45817.8613 & 10.88 & $-$0.42 & 11.84 & $+$0.80 & 30.70 & $+$8.90 & 30.22 & $-$0.32 & 11.52 & $-$0.68 &  6.00 & $+$2.94 \\ 
46461.9627 & 14.50 & $-$1.12 & 11.81 & $+$0.21 & 10.80 & $+$4.02 & 16.38 & $+$4.50 & 27.24 & $+$2.46 & 26.34 & $+$3.50 \\ 
46464.9225 & 11.00 & $+$3.89 & 14.55 & $+$2.84 & 15.95 & $+$2.93 &  9.33 & $-$2.71 & 25.86 & $-$1.14 & 24.62 & $+$2.11 \\ 
46485.9236 & 12.21 & $-$1.47 & 12.13 & $-$0.40 &  2.44 & $-$0.15 &  2.91 & $-$0.93 & 30.81 & $-$0.69 & 29.15 & $-$1.25 \\ 
46487.9269 & 13.30 & $-$0.21 & 13.01 & $+$0.39 & 32.38 & $-$1.36 & 32.42 & $+$0.66 & $-$2.20 & $+$0.32 &  1.08 & $+$1.59 \\ 
46489.9227 & 13.79 & $-$1.94 & 11.69 & $-$1.01 & 13.02 & $+$3.03 & 16.63 & $+$1.66 & 21.57 & $+$0.73 & 18.04 & $+$0.16 \\ 
46492.9182 & 12.72 & $+$0.18 & 11.60 & $-$1.24 & 16.14 & $+$1.22 & 13.09 & $-$2.83 & 18.49 & $-$0.47 & 16.24 & $-$0.40 \\ 
46511.9706 & 14.13 & $+$0.45 & 13.66 & $-$0.06 & 25.44 & $-$1.26 & 23.49 & $-$0.87 &  4.91 & $+$0.07 &  5.38 & $-$0.71 \\ 
46512.8394 & 14.77 & $+$1.37 & 13.92 & $+$0.16 &  0.42 & $+$0.02 & $-$0.22 & $+$0.88 & 34.80 & $+$0.67 & 31.27 & $-$2.85 \\ 
46513.8548 & 14.82 & $-$0.15 & 14.01 & $+$0.20 & 21.46 & $-$1.52 & 21.59 & $+$0.25 &  8.26 & $+$0.67 &  7.63 & $-$1.66 \\ 
46520.0076 & 14.34 & $-$0.71 & 14.01 & $-$0.10 & 27.21 & $-$1.20 & 25.54 & $-$1.79 &  1.23 & $-$0.12 &  2.75 & $+$0.49 \\ 
46520.8817 & 15.08 & $+$0.03 & 14.93 & $+$0.78 & 24.31 & $-$0.47 & 23.74 & $+$0.10 &  5.93 & $+$0.58 &  6.23 & $-$0.04 \\ 
46523.8260 & 14.33 & $+$0.45 & 13.75 & $-$0.54 & 25.28 & $-$4.27 & 22.28 & $-$1.90 &  1.32 & $-$0.32 &  4.19 & $-$1.29 \\ 
46537.9235 & 15.75 & $+$0.79 & 15.43 & $+$0.43 & 30.70 & $+$0.90 & 28.82 & $-$0.29 & $-$0.04 & $+$0.38 &  0.46 & $+$1.42 \\ 
46538.9682 & 12.11 & $+$5.30 & 12.37 & $-$2.68 & 18.73 & $+$3.45 & 18.90 & $+$5.80 & 20.65 & $-$3.31 & 17.43 & $+$0.80 \\ 
46539.8366 & 15.35 & $-$0.43 & 15.50 & $+$0.41 & $-$1.29 & $-$0.54 & $-$1.39 & $-$0.38 & 32.39 & $-$0.12 & 31.86 & $-$0.28 \\ 
46540.8298 & 15.00 & $-$0.40 & 15.75 & $+$0.61 & 28.30 & $-$0.80 & 27.17 & $-$0.81 & $-$0.60 & $-$0.52 &  1.34 & $+$1.24 \\ 
46540.9278 & 15.31 & $-$0.24 & 15.22 & $+$0.07 & 30.19 & $-$0.65 & 29.13 & $-$0.46 & $-$2.37 & $-$0.21 & $-$3.06 & $-$1.36 \\ 
46541.8861 & 14.86 & $+$1.45 & 14.03 & $-$1.17 & 16.54 & $-$3.58 & 13.08 & $-$1.63 & 11.59 & $-$0.71 & 17.12 & $+$2.48 \\ 
46565.7149 & 16.63 & $+$1.18 & 15.66 & $-$0.72 & 11.68 & $+$1.31 & 11.81 & $+$1.66 & 20.00 & $-$0.15 & 20.92 & $+$2.91 \\ 
\hline
\end{tabular}
\label{radial.velocities.tab}
\end{table*}

The elements for the spectroscopic orbits derived by the Tel Aviv team
are given in Table \ref{spectroscopic.elements.tab}.  For four of the
observed spectra the assignment of the three velocities to the three
stars in the system was swapped from the initial assignment made by
TODCOR in order to improve the overall $\chi^2$ of the solution, as
described above. We denote the orbit of Ba and Bb by B, and the orbit
of A and B by AB. The period, eccentricity, longitude of periastron,
and the time of periastron passage are denoted by $P, e, \omega$ and
$T$, respectively. The radial-velocity semi-amplitudes of A and B are
denoted by $K_{\rmn{A}}$ and $K_{\rmn{B}}$. Similar notations are used
for Ba and Bb.  We also report the light ratios derived using TODCOR
in Table \ref{spectroscopic.elements.tab}. The errors in the light
ratios were estimated using an analysis of the $\chi^2$ values near
the peak of the three-dimensional correlation surface.  These errors
should include the effects of photon noise, but do not take into
account possible systematic errors due to template mismatch. 

\begin{table*}
\caption{The spectroscopic elements for the inner and outer orbits.}
\begin{tabular}{lccc}
\hline
Element                                  &  Tel Aviv             & CfA             & Adopted \\
\hline
$P_{\rmn{B}}$ (days)                     &$2.965530 \pm 0.000011$&$2.965515 \pm 0.000018$& $2.965522 \pm 0.000014$ \\
$K_{\rmn{Ba}}$ (\kms)                    &$16.81 \pm 0.14$       &$17.20 \pm 0.17$       & $17.01 \pm 0.20$\\
$K_{\rmn{Bb}}$ (\kms)                    &$18.54 \pm 0.19$       &$19.00 \pm 0.21$       & $18.77 \pm 0.23$\\
$e_{\rmn{B}}$                            &$0.030 \pm 0.006$      &$0.021 \pm 0.007$      & $0.026 \pm 0.007$\\
$\omega_{\rmn{B}} (\degr)$               &$170   \pm 12  $       &$162 \pm 21$           & $166 \pm 16$\\
$T_{\rmn{B}}$ (HJD-2440000)              &$7337.3 \pm 0.1$       &$7337.23 \pm 0.18$     & $7337.3 \pm 0.14$  \\
$a_{\rmn{Ba}} \sin i_{\rmn{B}}$ (Gm)     &$0.685 \pm 0.006$      &$0.701 \pm 0.007$      & $0.693 \pm 0.008$ \\
$a_{\rmn{Bb}} \sin i_{\rmn{B}}$ (GM)     &$0.756 \pm 0.008$      &$0.775 \pm 0.009$      & $0.766 \pm 0.009$ \\
$M_{\rmn{Ba}} \sin^3 i_{\rmn{B}} (\msun)$&$0.00713 \pm 0.00016$  &$0.00765 \pm 0.00019$  & $0.00739 \pm 0.00026$\\
$M_{\rmn{Bb}} \sin^3 i_{\rmn{B}} (\msun)$&$0.00646 \pm 0.00013$  &$0.00692 \pm 0.00016$  & $0.00669 \pm 0.00023$\\
$q_{\rmn{B}} \equiv M_{\rmn{Bb}}/M_{\rmn{Ba}}$&$0.906 \pm 0.012$&$0.905 \pm 0.014$&$0.906 \pm 0.015$\\
$\sigma_{\rmn{Ba}}$ (\kms)               &$1.35$                 &$1.7$                  & \\
$\sigma_{\rmn{Bb}}$ (\kms)               &$1.88$                 &$2.1$                  & \\
\\
$P_{\rmn{AB}}$ (days)                    &$625.9 \pm 1.1$        &$625.7 \pm 1.6$        & $625.8 \pm 1.3$ \\
$K_{\rmn{A}}$ (\kms)                     &$4.81 \pm 0.08$        &$4.38 \pm 0.11$        & $4.60 \pm 0.22$\\
$K_{\rmn{B}}$ (\kms)                     &$3.22 \pm 0.10$        &$2.67 \pm 0.12$        & $2.95 \pm 0.28$ \\
$e_{\rm{AB}}$                            &$0.025 \pm 0.016$      &$0.080 \pm 0.024$      & $0.053 \pm 0.028$ \\
$\omega_{\rmn{AB}} (\degr)$              &$311  \pm 36$          &$285 \pm 17$           & $298 \pm 27$ \\
$T_{\rmn{AB}}$ (HJD-2440000)             &$7232 \pm 62$          &$7185 \pm 28$          & $7208 \pm 45$ \\
$a_{\rmn{A}} \sin i_{\rmn{AB}}$ (Gm)     &$41.42 \pm 0.67$       &$37.53 \pm 0.96$       & $39.5 \pm 2.0$ \\
$a_{\rmn{B}} \sin i_{\rmn{AB}}$ (Gm)     &$27.72 \pm 0.87$       &$22.93 \pm 1.08$       & $25.3 \pm 2.4$ \\
$M_{\rmn{A}} \sin^3 i_{\rmn{AB}} (\msun)$&$0.01348 \pm 0.00080$  &$0.00853 \pm 0.00075$  & $0.0110 \pm 0.0025$ \\
$M_{\rmn{B}} \sin^3 i_{\rmn{AB}} (\msun)$&$0.02014 \pm 0.00089$  &$0.0140 \pm 0.0009$    & $0.0171 \pm 0.0031$ \\
$q_{\rmn{AB}} \equiv M_{\rmn{B}}/M_{\rmn{A}}$&$1.494 \pm 0.048$&$1.636 \pm 0.087$& $1.56 \pm 0.17$\\
$\sigma_{\rmn{A}}$ (\kms)                &$0.84$                 &$1.2$ & \\
\\
$\gamma$ (\kms)                          &$14.89 \pm 0.05$       &$15.30 \pm 0.06$       &$15.10 \pm 0.21$\\
$N_{\rmn{obs}}$                          &$200 \times 3$         &$200 \times 3$         &\\
$\alpha$                                 &$0.56 \pm 0.06$        &$0.571 \pm 0.008$      &$ 0.566 \pm 0.034$\\
$\beta$                                  &$0.36 \pm 0.04$        &$0.355 \pm 0.007$      &$ 0.358 \pm 0.024$\\
\hline
\end{tabular}
\label{spectroscopic.elements.tab}
\end{table*}

%
 
\begin{figure}
\def\epsfsize#1#2{0.5#1}
\epsfbox{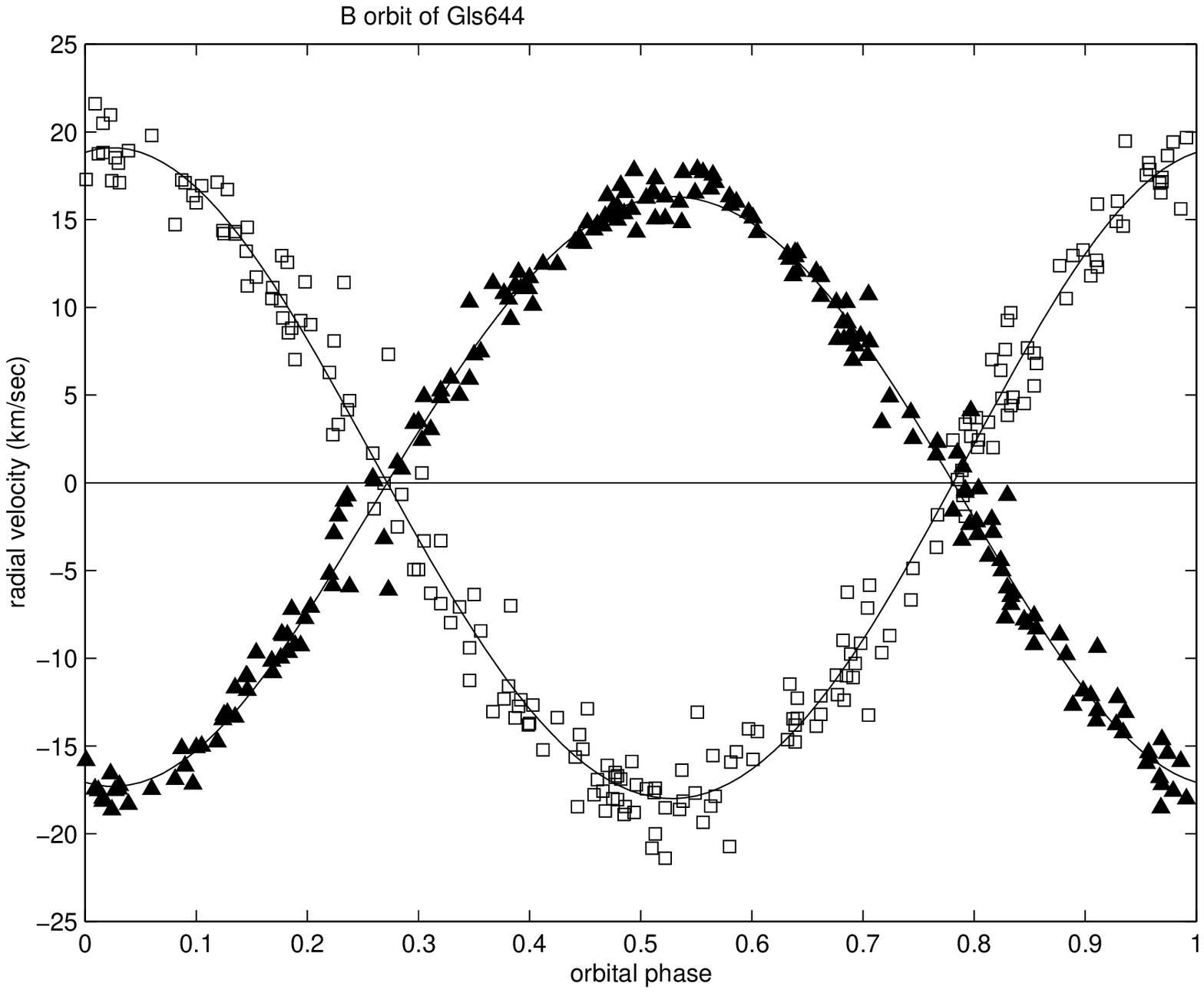}
\vspace{0.1in}
\caption{The radial-velocity curve for the inner orbit. The motion of
the center of mass of the B system around A has been removed from the
velocities plotted for Ba and Bb, represented by filled triangles and
open squares, respectively.}
\label{spectroscopic.inner}
\end{figure}

\begin{figure}
\def\epsfsize#1#2{0.5#1}
\epsfbox{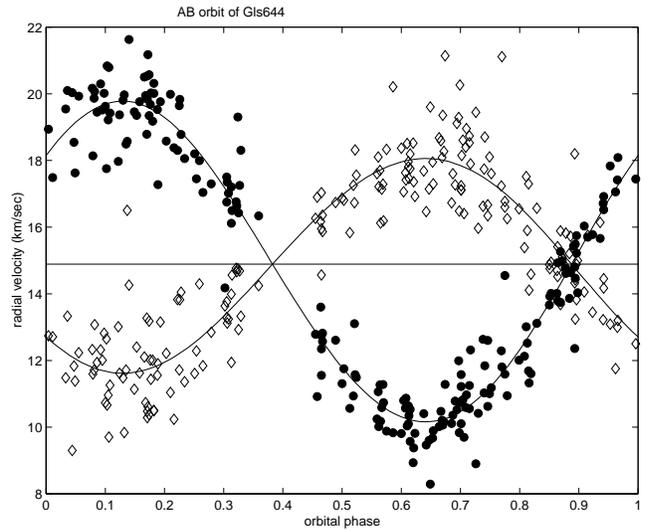}
\vspace{0.1in}
\caption{The radial-velocity curve for the outer orbit.  The
velocities observed for A are plotted as filled circles, and open
diamonds are used for the average velocity of Ba and Bb, weighted
according to their mass ratio.}
\label{spectroscopic.outer}
\end{figure}

Figure \ref{spectroscopic.inner} depicts the radial-velocity curve for
the inner orbit of the close spectroscopic pair Ba and Bb according to
the Tel Aviv solution.  The velocities observed for Ba and Bb are
plotted with the motion of the center of mass of the B system around A
removed.  The radial-velocity curve for the outer orbit of the wide
visual pair A and B is presented in Figure
\ref{spectroscopic.outer}. The velocity plotted for B is the
average of the velocities observed for Ba and Bb, weighted according
to their mass ratio --- $K_{\rmn{Ba}}:K_{\rmn{Bb}}$.

\subsubsection{The CfA analysis}

\begin{figure*}
\vspace{-0.1in}
\def\epsfsize#1#2{0.75#1}
\epsfbox{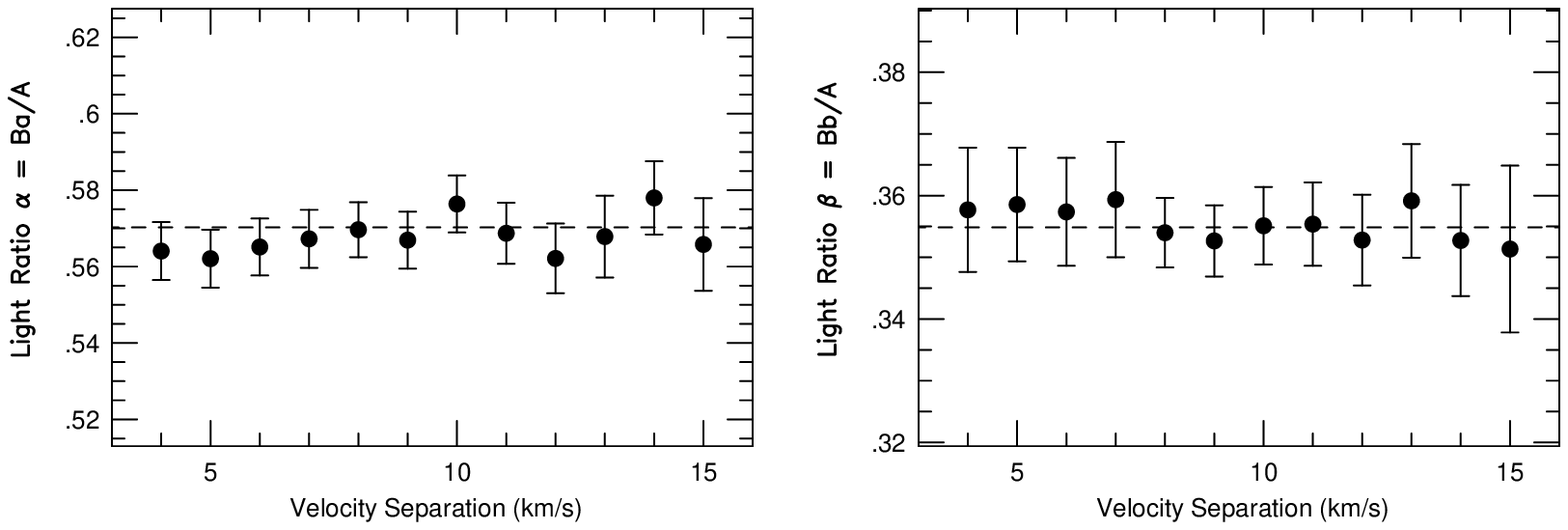}
\vspace{-5.7in}
\caption{The light ratios $\alpha$ and $\beta$ from the CfA analysis
as a function of the minimum velocity separation between any two
components of the Gliese 644 system.}
\label{alphabeta}
\end{figure*}

At CfA the standard approach has been to select the optimum templates
and determine the light ratio for a double-lined binary by working with
a subset of the observed spectra, just the ones where the lines of the
two stars were well resolved.  This did not work well for Gliese 644,
because the lines were resolved in only a few spectra. We addressed this
problem by sorting the spectra into an order ranked according to the
minimum velocity separation between any two of the components in the
system.  Then we calculated the average light ratios for all the spectra
with velocity separations larger than some specified minimum value.
Plots of the average light ratio versus minimum separation were used to
select the light ratios for the next iteration.  Figure \ref{alphabeta}
shows these $\alpha$ and $\beta$ plots for the final CfA solution.  The
fact that these plots are quite flat is an indication that we have
arrived at a robust solution where the light ratio is essentially
independent of blending, down to a minimum separation of 5~\kms.  For
each point plotted in Figure \ref{alphabeta} the error bars are the
standard deviation of the mean value of the light ratio.  The scatter of
the points is similar to the error bars, suggesting that these errors
are reasonable estimates of the internal precision. The errors quoted
for the CfA light ratios in Table \ref{spectroscopic.elements.tab} are
estimated from these error bars and scatter, and are much smaller than
the corresponding errors reported for the Tel Aviv analysis. 

For each iteration in the CfA analysis we used the orbital solution from
the previous iteration to provide TODCOR with initial values for the
velocities of the three stars.  We then allowed TODCOR to search for the
nearest peak in correlation space.  This procedure should eliminate or
at least minimize the cases where TODCOR finds the wrong correlation
peak and assigns an incorrect velocity to one or more of the stars.
Therefore, for the CfA analysis we chose not to swap any of TODCOR's
velocity assignments. The orbital solution was carried out with an
independent code that solves simultaneously for all elements of the
inner and outer orbits, by computing differential corrections to a set
of initial elements.  The results are reported in Table
\ref{spectroscopic.elements.tab}, and the (O-C) residuals of the
individual velocities from this solution are listed in Table
\ref{radial.velocities.tab}.

\subsection{The adopted spectroscopic orbits}

The agreement between the Tel Aviv and CfA analyses of the spectroscopic
orbits is satisfactory for all of the orbital parameters except the
velocity amplitudes for the outer orbit, $K_{\rmn{A}}$ and
$K_{\rmn{B}}$, and the center-of-mass velocity, $\gamma$, where the
differences exceed three times the combined internal error
estimates. These differences lead to a relatively large difference in
the mass ratio $q_{\rmn{AB}}= K_{\rmn{A}}:K_{\rmn{B}}$, which we use to
divide up the total mass from the astrometric orbit in the next section.
Apparently there is a systematic error lurking in one or both of our
analyses, but we have not been able to identify the culprit.  Therefore
we have chosen to adopt final orbital parameters that are simply the
average of the values from the two solutions, as reported in Table
\ref{spectroscopic.elements.tab}.  For the final errors we adopted half
the difference between the two solutions, or the average of the two
internal error estimates, whichever was larger.

The agreement between the astrometric orbit and our adopted
spectroscopic orbit is well within the errors, which are much smaller
for the astrometric solution.  By convention, the longitude of
periastron for an astrometric orbit refers to the secondary, but for a
spectroscopic orbit it refers to the primary.  After this difference
of $180\degr$ is taken into account, the values for $\omega$ agree
well.

\section{THE MASSES AND ORBITAL INCLINATIONS OF GLIESE 644}

\subsection{The masses of the three stars}

The combination of spectroscopic orbits for the inner and outer
binaries together with an astrometric orbit for the outer visual
binary enables us to derive masses for all three components of Gliese
644. The total mass of the system can be derived from the parallax
together with the dynamical quantity $(a^3/P^2)$ provided by the
astrometric solution. The individual masses can then be deduced using
the two mass ratios derived from the spectroscopic solutions. 

\begin{figure}
\vspace{-0.7in}
\def\epsfsize#1#2{0.4#1}
\epsfbox{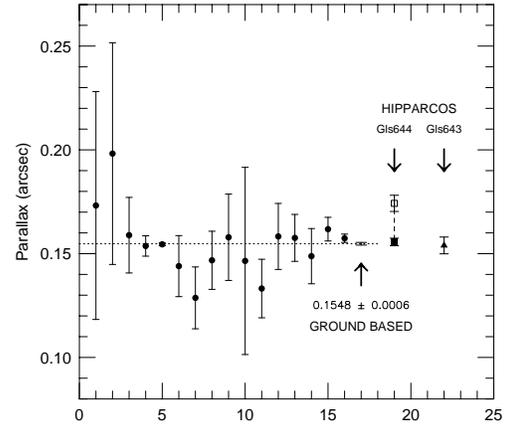}
\vspace{-0.6in}
\caption{Parallax determinations for Gliese 644.  The individual
ground-based results are plotted as circles. The weighted average is
indicated by the dotted line. The arrow points to the corresponding
error bar. S\"oderhjelm's (1999)
redetermination for Gliese 644 is plotted as a filled square,
replacing the Hipparcos value plotted as an open square.}
\label{parallax}
\end{figure}

For the parallax we note that the Fourth Edition of the General Catalog of
Trigonometric Parallaxes (van Altena, Lee \& Hoffleit 1995) gives the
weighted average of 16 determinations as $\pi_{\rm trig} =
0\th\farcs1548 \pm 0\th\farcs0006$. The Hipparcos catalog (ESA
1997) lists a considerably larger value ($\pi_{\rm HIP} = 0\th\farcs1742
\pm 0\th\farcs0039$), which is almost certainly affected by the 1.7-yr
orbital motion of the visual pair.  S\"oderhjelm (1999) took this
orbital motion into account in his reanalysis of the Hipparcos data and
derived $\pi = 0\th\farcs1556 \pm 0\th\farcs0018$, in excellent
agreement with the ground-based value.  These determinations are plotted
in Figure \ref{parallax}, including the 16 individual ground-based
values.  A valuable check is provided by Gliese 643, which is a
common-proper-motion companion to Gliese 644 and is therefore expected
to be at the same distance (see Section 7). The value of its parallax as
measured by Hipparcos is $\pi_{\rm HIP} = 0\th\farcs1540 \pm
0\th\farcs0040$, again in good agreement with the ground-based value
for Gliese 644. We adopt the ground-based value for the
parallax of Gliese 644 because it has the smallest estimated error.

Combining $(a^3/P^2)$ from our adopted astrometric orbit with the
parallax we get that the total mass is
\begin{equation}
M_{\rmn{tot}} = \frac{(a^3/P^2)}{\pi^3} = 1.049 \pm 0.020\ (1.9\%)\ 
\msun\,.
\end{equation}

An alternative way to derive the semi-major axis of the wide binary, and
consequently the total mass of the system, is to use the radial-velocity
amplitudes derived from the spectroscopy together with the inclination
angle derived from the astrometry.  This yields $M_{\rmn{tot}}=1.17 \pm
0.40 \msun$.  Note that the error for this approach is an order of
magnitude larger than the one corresponding to the mass derived from the
astrometry alone, and the total masses from the two approaches agree
within this large error.

The mass ratios given by our adopted spectroscopic orbits are
$q_{\rmn{B}}= M_{\rmn{Bb}}:M_{\rmn{Ba}} = 0.906 \pm 0.015$ for
the inner orbit and $q_{\rmn{AB}}= M_{\rmn{B}}:M_{\rmn{A}} =
1.56 \pm 0.17$ for the outer orbit.  
Combining these mass ratios
with the definitions $M_{\rmn{tot}} = M_{\rmn{A}} + M_{\rmn{B}}$ and
$M_{\rmn{B}} = M_{\rmn{Ba}} + M_{\rmn{Bb}}$, we conclude that

\begin{equation}
M_{\rmn{A}} = \frac{1}{(1+q_{\rmn{AB}})} M_{\rmn{tot}} 
= 0.410 \pm 0.028\ (6.9\%)\ \msun\,,
\end{equation}
\begin{equation}
M_{\rmn{Ba}} = \frac{q_{\rmn{AB}}}
{(1+q_{\rmn{B}})(1+q_{\rmn{AB}})} M_{\rmn{tot}}
= 0.336 \pm 0.016\ (4.7\%)\ \msun\,,
\end{equation}
\begin{equation}
M_{\rmn{Bb}} = \frac{q_{\rmn{B}}q_{\rmn{AB}}}
{(1+q_{\rmn{B}})(1+q_{\rmn{AB}})} M_{\rmn{tot}}
= 0.304 \pm 0.014\ (4.7\%)\ \msun. 
\end{equation}

These values are similar to the ones found by S\"oderhjelm
(1999), who derived $M_{\rmn{A}} = 0.41 \pm 0.04$ and $M_{\rmn{B}} =
0.66\pm 0.06 \msun$, and to the ones derived 
by S\'egransan et al. (2001), who obtained masses of
$0.4155 \pm 0.0057$, 
$0.3466 \pm 0.0047$ and 
$0.3143 \pm 0.0040$, for
the masses of $A$, $Ba$ and $Bb$, respectively. 

\subsection{The orbital inclinations}

We can now go back and calculate the inclinations of the two orbits
relative to our line of sight, using the spectroscopic values for $M
\sin ^3i$ adopted in Table \ref{spectroscopic.elements.tab}, together
with the individual masses derived in the previous section. We get
\begin{equation}
i_{AB}=17\fdg4 \pm 1\fdg4\ {\rm or} \ 162\fdg6 \pm1\fdg4, \ \ \ 
i_B = 16\fdg3 \pm 0\fdg3 \ {\rm or} \ 163\fdg7 \pm0\fdg3 \,.
\end{equation}
The ambiguity comes from the fact that the radial-velocity data can
not specify whether the orbital motions are retrograde or direct.  The
inclination of the outer orbit from the astrometric solution, $i_{AB}
= 163\fdg1 \pm 1\fdg6$, is close to the retrograde value from the
spectroscopic orbit.

The inclinations of the two orbits do
not provide enough information to derive the relative angle between
the two orbits, $\phi$, which can be written as (Mazeh \&
Shaham 1976):
\begin{equation}
\cos \phi = \cos i_{\rmn{B}} \cos i_{\rmn{AB}} 
          + \sin i_{\rmn{B}} \sin i_{\rmn{AB}} 
            \cos (\Omega_{\rmn{B}} - \Omega_{\rmn{AB}})\,,
\end{equation}
where $\Omega_{\rmn{B}}$ and $\Omega_{\rmn{AB}}$ are the position
angles of the line of nodes of the inner and outer orbits,
respectively.  Since $\Omega_{\rmn{B}}$ is unknown, we can only limit
$\phi$ (Batten 1973):
\begin{equation}
i_{\rmn{A}} - i_{\rmn{AB}} \leq \phi \leq i_{\rmn{A}} + i_{\rmn{AB}},
\end{equation}
which in our case results in
\begin{equation}
1\fdg1 \pm 1\fdg4\leq \phi \leq 33\fdg7 \pm 1\fdg4\,.
\end{equation}

Equation (8) implies that the relative angle between the two orbits
could be as large as $34 \degr$. However, to make the two inclinations
come out so close requires a special orientation.  It is more natural to
suppose that the two orbits really are almost coplanar.

\section{PHOTOMETRY}

The Hipparcos Catalogue lists $V = 9.02$ mag for the total combined
light of Gliese 644, while SIMBAD cites 6 determinations that lead to $V
= 9.02 \pm 0.01$ mag.  To divide the total light between the three
components we use the adopted values of the spectroscopic light ratios
at 5187~\AA\ derived by TODCOR, $\alpha = 0.569 \pm 0.034$ and $0.358
\pm 0.024$.  After the application of small corrections to move these
light ratios to 5550~\AA, the value we adopted for the effective
wavelength of the $V$ band, these ratios become $0.576 \pm 0.034$ and
$0.364 \pm 0.024$. Dividing the $V$ brightness with these light
ratios, and applying our adopted parallax of $\pi = 0\th\farcs1548 \pm
0\th\farcs0006$, we find that 
$M_V\rmn{(A)} = 10.69 \pm 0.02$,
$M_V\rmn{(Ba)} = 11.29 \pm 0.05$, and  
$M_V\rmn{(Bb)} = 11.79 \pm 0.05$ mag.

As a check on the spectroscopic light ratios derived with TODCOR, we
calculate the total light that those ratios predict for the B
subsystem and compare it with the magnitude difference between A
and B found by others.  The simple average of the 70 visual magnitude
differences reported in the Washington Double Star Catalog (Worley \&
Douglass 1996) gives $0.103 \pm 0.014$.  S\"oderhjelm (1999) reports a
difference in the Hipparcos magnitudes of 0.11 based on his reanalysis
of the Hipparcos data. These magnitude differences are both consistent
with our result that the total light of B is fainter than A by $0.07
\pm 0.05$ mag in the $V$ band. 

The three new speckle observations reported in this paper provide
infrared magnitude differences, and show that B is {\it brighter} than
A at $J$, $H$, and $K$ by $0.51 \pm 0.01$, $0.56 \pm 0.02$, and $0.66
\pm 0.06$ mag, respectively.  The fact that B is fainter than A in the
visual but brighter in the infrared is qualitatively consistent with
the fact that the two components of B are cooler than A. 

\begin{figure}
\vspace{-0.0in}
\def\epsfsize#1#2{0.4#1}
\epsfbox{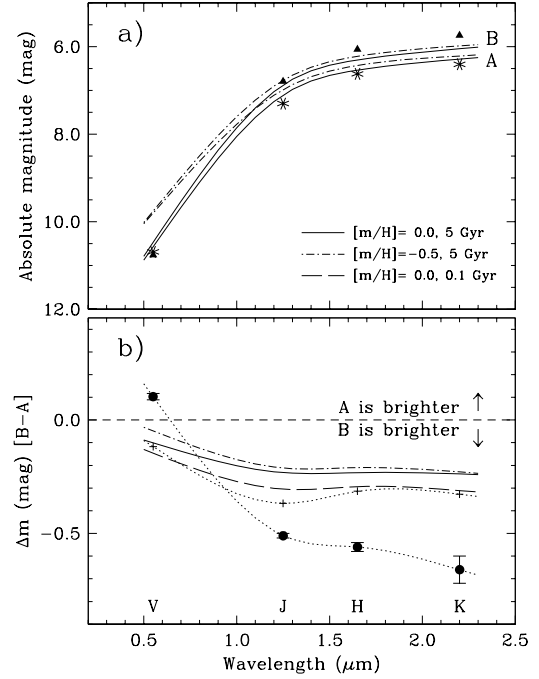} 
\vspace{-0.2in}
\caption{Absolute magnitudes (panel a) and magnitude differences (panel
b) for Gliese 644 A and B. In panel a the observed A magnitudes are denoted by
stars, and those of B by triangles.  The predicted absolute magnitudes
for A and B from 5 Gyr solar-metallicity models are shown as solid
lines, and from [Fe/H] = $-0.5$ metal-poor models as dash-dotted lines.
In panel b a 0.1 Gyr
solar-metallicity prediction is plotted as a dashed line.
The plus signs present the magnitude differences predicted
by the Henry \& McCarthy (1993) empirical mass-luminosity relations.
}
\label{absolute.magnitudes}
\end{figure}

However, when we use our derived masses to predict the magnitudes
expected for A and B in detail, a problem emerges. This is illustrated
in Figure \ref{absolute.magnitudes}a, where we plot the observed
absolute magnitudes for A (tesselated stars) and B (filled triangles) at
$V$, $J$, $H$, and $K$, using the apparent magnitudes for Gliese 644
(all three stars together) of $J=5.37$, $H=4.67$, and $K=4.38$. We used
Leggett (1992) observations and applied small corrections to convert
them to the Johnson system.  We also plot the predicted absolute
magnitudes as a function of wavelength from 5 Gyr solar-metallicity
models (solid lines, Baraffe et al.\ 1998) and [Fe/H] = $-0.5$
metal-poor models (dash-dotted lines, Baraffe et al.\ 1997).  In both
cases the lower line of each pair corresponds to the prediction for A,
and the upper line for B. 
The infrared absolute magnitudes predicted for A are slightly
brighter than observed, while the corresponding predictions for B are
typically too faint. In the visual band, the solar-metallicity model
is slightly too bright for both stars.

 These
discrepancies become more obvious when we plot the magnitude
differences (B$-$A), as shown in panel b.  The 5 Gyr solar-metallicity
and [Fe/H] = $-0.5$ metal-poor model predictions are again plotted as
solid and dash-dotted lines, respectively, while a 0.1 Gyr
solar-metallicity prediction is plotted as a dashed line.  Although
metallicity and age both have some effect on the predicted magnitude
differences, the effect is much smaller than the discrepancy with the
observed magnitude differences.  To show that the models appear to be
reliable, we plot (as plus signs) the magnitude differences predicted
by the Henry \& McCarthy (1993) empirical mass-luminosity relations.
We are puzzled by these discrepancies and see no easy explanation. 

\section{THE MASS-LUMINOSITY RELATION AND GLIESE 644}

The individual masses that we have derived for the three M dwarfs in
the Gliese 644 system have formal errors of 5 to 7 \%, which puts these
results into a category with only a dozen or so stars that have
masses smaller than 0.6 \msun\ and mass uncertainties similar or better
than ours.  At
this level of accuracy we can begin to make meaningful tests of the
theoretical models for low-mass stars. 

It has been traditional to assess the mass-luminosity relation using
the absolute $V$ magnitude versus mass diagram.  We can plot all three
stars in the Gliese 644 system on this diagram, because TODCOR
has provided us with the spectroscopic light ratios needed to divide
up the light between the two components of the unresolved inner binary
at a wavelength close to the V band.
Thus we use the $M_V$ versus mass diagram to assess the
present status of the confrontation between the observations and the
Baraffe et al.\ (1997, 1998) theoretical isochrones for low-mass stars.

\begin{table*} \caption{Lower main-sequence stars with mass and V
       absolute magnitude better than 5\%.}
\begin{tabular}{llll}
\hline
Star        &       Mv         &     Mass             &  Ref.\\
\hline
Gls 791.2 A &13.38 $\pm$ 0.03 & 0.2866 $\pm$ 0.0061 & Benedict et al. 2000\\ 
Gls 791.2 B &16.65 $\pm$ 0.10 & 0.1258 $\pm$ 0.0029 &        \\ 
\\
Gls 860 A   &11.81 $\pm$ 0.07 & 0.257\phantom{0}  $\pm$ 0.011  & Henry \& McCarthy 1993\\
Gls 860 B   &13.39 $\pm$ 0.06 & 0.172\phantom{0}  $\pm$ 0.008  & Henry et al. 1999\\
\\
GJ 2069 A   &11.7\phantom{0}  $\pm$ 0.2  & 0.4329 $\pm$ 0.0018 & Delfosse et al. 1999a\\
GJ 2069 B   &12.45 $\pm$ 0.2  & 0.3975 $\pm$ 0.0015 &        \\
\\
Gls 866 A   &15.34 $\pm$ 0.14 & 0.1216 $\pm$ 0.0029 & Delfosse et al. 1999b\\
Gls 866 B   &15.58 $\pm$ 0.07 & 0.1161 $\pm$ 0.0029 &        \\
Gls 866 C   &17.34 $\pm$ 0.45 & 0.0957 $\pm$ 0.0023 &        \\
\\
YY Gem A    &\phantom{1}8.99 $\pm$ 0.08 & 0.588\phantom{0}  $\pm$ 0.022  & Andersen 1991\\
YY Gem B    &\phantom{1}8.99 $\pm$ 0.08 & 0.588\phantom{0}  $\pm$ 0.022  &       \\
\\
CM Dra A    &12.75 $\pm$ 0.04 & 0.2307 $\pm$ 0.0010 & Metcalfe et al. 1996\\
CM Dra B    &12.90 $\pm$ 0.04 & 0.2136 $\pm$ 0.0010 &   \\
\\
Gls 644 A   &10.69 $\pm$ 0.02 & 0.410 $\pm$ 0.028 &    This paper\\
Gls 644 Ba  &11.29 $\pm$ 0.05 & 0.336 $\pm$ 0.016 &    \\
Gls 644 Bb  &11.79 $\pm$ 0.05 & 0.304 $\pm$ 0.014 &    \\
\hline
\end{tabular}
\label{lower_main_sequence.tab}
\end{table*}

\begin{figure}
\vspace{-0.9in}
\def\epsfsize#1#2{0.4#1}
\epsfbox{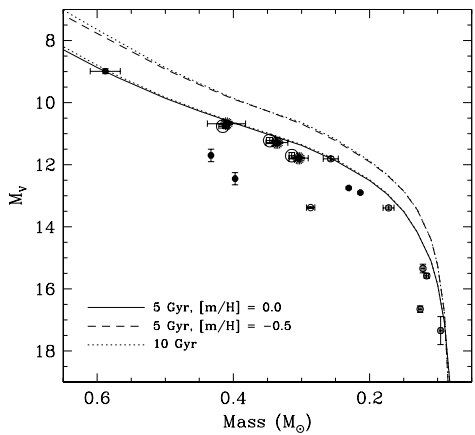}
\vspace{-0.9in}
\caption{The mass-luminosity relation for stars on the lower main
sequence with mass determinations more accurate than 5\%.  The three
components of Gliese 644 are plotted as tesselated stars.  For
comparison we also show the determinations for Gliese 644 by S\'egransan
et al. (2001). The components of the three eclipsing binaries YY Gem, GJ
2069A, and CM Dra are plotted as filled circles, and the components of
the astrometric systems Gliese 860, Gliese 866, and Gliese 791.2 as open
circles.  The solid line is the solar-metallicity 5 Gyr isochrone, and
the dashed line is the metal-poor 5~Gyr isochrone with [Fe/H] = $-0.5$.
The corresponding 10 Gyr isochrones are plotted as dotted lines.  }
\label{mass.luminosity}
\end{figure}

In Table 6 we list the masses and absolute visual magnitudes obtained
here for Gliese 644, as well as data for other stars below 0.6 \msun\
with uncertainties in the masses smaller than 5\%, which we take as a
benchmark for comparison. These measurements are shown in Figure
\ref{mass.luminosity}. We did not list and plot Gliese 570
BC (Forveille et al.\ 1999) because that close system has not been
resolved in the visual, and therefore observed $V$ magnitudes are not
available for the two components.

In the figure we plot solar-metallicity 5 Gyr isochrone from Baraffe et
al.\ (1998), and metal-poor 5 Gyr isochrone with [Fe/H] = $-0.5$ from
Baraffe et al.\ (1997).  The corresponding 10 Gyr isochrones are also
plotted.  As can be seen from the figure, some of the observed points
are far off from both theoretical lines.

Forveille et al.\ (1999) suggest that the Baraffe et al.\ models (1997,
1998) may be missing some opacity sources, and that the
solar-metallicity isochrones should be lower than they appear in Figure
\ref{mass.luminosity}.  This might produce better agreement with the
observed results for CM Dra, Gliese 866 and 860, in which case the good
agreement of our three points for Gliese 644 might imply that the system
is slightly metal poor and/or very young.  Can the much more serious
discrepancies for GJ 2069A, as suggested by Delfosse et al.\ (1999a),
and for Gliese 791.2, be explained by extreme metal richness?  The
answer to this question will have to wait for the calculation of
theoretical isochrones for super-metal-rich models and for the
observational determination of accurate metallicities for these stars.
In the meantime we must conclude that the mass-luminosity relation near
the bottom of the main sequence is still not well understood, at least
in the visual band.  Note that the above dicussion did not take into
account the possiblity of stellar spots, which could affect the stellar
luminosity. In the future, it might be better to move to infrared
magnitudes, where the effects of stellar spots on the photometry are
less severe.

Figure \ref{mass.luminosity} shows that changing the stellar age from 5
to 10 Gyr has no significant effect on the theoretical isochrones for
stars less massive than about 0.5 \msun.  However, changing the
metallicity from [Fe/H] = 0.0 to $-0.5$ has a large effect.  To allow a
meaningful confrontation between the models and the observations in the
$M_V$ versus mass diagram, it is necessary to have accurate
metallicities for the stars involved, good to 0.1 in [Fe/H] or better.
Otherwise, the metallicity uncertainties are likely to dominate the
uncertainties from the photometry, parallaxes, and masses.
Unfortunately, accurate metallicity determinations are still beyond the
state of the art for stars near the bottom of the main sequence.  Of all
the systems plotted in Figure \ref{mass.luminosity}, for example, only
for CM~Dra has the metallicity been analyzed in some detail, with
ambiguous results (Viti et al.\ 1998).  They find that CM~Dra may be
somewhat metal poor, which would be consistent with its unusually high
space motion, or it may have almost solar metallicity, but it is
unlikely to be metal rich.

Actually, it is possible that metallicity alone is not enough to
understand the mass-luminosity relation.  For example, it may turn out
that it is not valid to assume that the helium abundance scales with the
metallicity, in which case it will also be necessary to determine the
bulk helium abundance. Stellar age is certainly another factor that can
change the stellar 
luminosity.  We expect therefore that studies of stars in clusters will
(continue to) be important for progress in testing stellar models,
because additional information about the age and metallicity can be
derived for cluster stars.
 
\section{THE DISTANT COMPANIONS OF GLIESE 644}

In this section we discuss two additional stars associated with the
Gliese~644 triple system. One of them is Gliese~643 (=Wolf 629,
$\alpha = 16$:55:25.26, $\delta = -8$:19:21.3 [J2000], $V=11.74$ mag), at
a projected separation of about $70\arcsec$ from Gliese~644.  The
other companion, at a separation of $220\arcsec$, is the faint star
vB~8 (=Gliese 644C, $\alpha = 16$:55:35.74, $\delta = -08$:23:36.0
[J2000], $V=16.80$ mag). The common proper motion of Gliese 644 and
643 (Wolf 1919) and their similar parallax (e.g., ESA 1997) strongly
indicate that they are indeed physically connected. Van Biesbroeck
(1961) found that vB~8 also shares with Gliese 644 the same proper
motion, attesting to its physical association with the system. 

vB 8 (spectral type M7.0 V) and vB 10 (M8.0 V), a companion to Gliese
752 also found by van Biesbroeck, were long considered to represent the
bottom of the stellar main sequence.  Although many cooler and less
luminous objects are now known, vB 8 is still of interest, because its
mass must be very near the substellar limit of 0.08 \msun.  Using the
updated mass-luminosity relation of Henry et al.\ (1999), we derive from
the $M_V$ values of Gliese~643 and vB~8 (12.69 and 17.75 mag) mass
estimates of 0.19 and 0.08 \msun, respectively.

Joy (1947, see Abt 1973) obtained a few spectra of Gliese~643 and
reported that its radial velocity was variable.  Since that time
Gliese~643 has usually been reported in the literature as a
spectroscopic binary (e.g., Eggen 1978, Johnson 1987). To follow its
orbital motion, we secured 83 spectra of Gliese~643, spread over 5843
days, but could {\it not} find any significant radial-velocity
variation.  Our velocities, derived using Gliese 725B as the template,
have an r.m.s of 0.63 \kms, slightly larger than our typical errors
for M dwarfs, and the $\chi^2$ probability is small, but we can not
find any periodicity in the velocities or any orbital solution that
pass our usual tests for being significant.  If Gliese 643 is a
spectroscopic binary, the orbital amplitude must be less than about
0.5 \kms.  Our individual velocities for Gliese 643 are reported in
Table \ref{gliese643.velocities}. 

\begin{table}
\caption{Radial velocities and internal error estimates (km s$^{-1}$)
for Gliese 643 (first 20 lines)}
\begin{tabular}{ccc}
\hline
HJD & $V_{\rm{r}}$ & $\sigma_{\rm{int}}$ \\
\hline
2445780.9993 & 16.44 & $\pm 0.41$ \\
2446511.9753 & 15.84 & $\pm 0.26$ \\
2446512.8511 & 15.82 & $\pm 0.45$ \\
2446513.8675 & 15.87 & $\pm 0.59$ \\
2446520.0150 & 15.94 & $\pm 0.33$ \\
2446520.8916 & 16.00 & $\pm 0.85$ \\
2446523.8416 & 15.78 & $\pm 0.39$ \\
2446537.9360 & 15.73 & $\pm 0.20$ \\
2446538.9771 & 16.22 & $\pm 0.26$ \\
2446539.8486 & 15.79 & $\pm 0.60$ \\
2446540.8417 & 15.03 & $\pm 0.32$ \\
2446540.9364 & 15.83 & $\pm 0.21$ \\
2446541.8916 & 15.17 & $\pm 0.22$ \\
2446565.7301 & 15.42 & $\pm 0.43$ \\
2446568.7910 & 15.97 & $\pm 0.34$ \\
2446569.6356 & 15.75 & $\pm 0.79$ \\
2446569.7865 & 15.58 & $\pm 0.53$ \\
2446569.7964 & 16.27 & $\pm 0.59$ \\
2446608.6348 & 15.54 & $\pm 0.94$ \\
2446612.7619 & 14.80 & $\pm 0.69$ \\
\hline
\end{tabular}
\label{gliese643.velocities}
\end{table}

The mean radial velocity we find for Gliese 643 is $15.81 \pm 0.07$
\kms on the CfA system, i.e.\ with the same zero point as the CfA
velocities for Gliese 644.  This is very close to the center-of-mass
velocity of Gliese 644, $\gamma = 15.10 \pm 0.21$.
The projected separation between Gliese 644 and 643 is about 450 AU,
which corresponds to an orbital velocity of about 1.5~\kms\ for a
circular orbit.  The observed velocity difference is only 0.7 \kms,
consistent with the interpretation that Gliese 643 is gravitationally
bound in an orbit with Gliese 644.

We have also secured one low S/N spectrum of vB 8 with the Digital
Speedometer on the MMT and found its radial velocity to be $15.68 \pm
1.09$ \kms, consistent with the interpretation that vB 8 is bound to
the Gliese 643/644 system in a hierarchical quintuple configuration.
In this picture vB~8 moves around the center of mass of the Gliese
644/643 system. 

The projected separation between vB 8 and Gliese 644 is only three
times larger than the projected separation between Gliese 643 and
Gliese 644. Such a small ratio usually renders triple systems
dynamically unstable (e.g., Eggelton \& Kiseleva 1995, Kiseleva et al.
1995). Therefore, we suggest that the actual separation between vB 8
and Gliese 644 is significantly larger than its projected separation,
by at least a factor of two. 

\section{ON THE FORMATION AND EVOLUTION OF THE SYSTEM}

Our picture of the Gliese 644/643/vB~8 system can be summarized as
follows:

\begin{itemize}
\item All five stars reside in an hierarchical system. The orbital
sizes are of the order of 2000 (projected), 500 (projected), 1 and 
0.05 AU for vB~8, Gliese 643, Gliese 644AB, and Gliese 644B, respectively. 

\item The masses of vB~8, Gliese 643, Gliese 644A/Ba/Bb are of the order
0.1, 0.2, 0.4, 0.3, 0.3 \msun, respectively.

\item The innermost two orbits, in the Gliese 644 system itself, are
probably coplanar.

\end{itemize}

In this section we briefly discuss two extremely simplistic scenarios
for the formation and evolution of the system.  First we consider a
scenario in which the quintuple was formed by a sequence of
fragmentation events during the collapse of a molecular cloud core
(e.g., Burkert \& Bodenheimer 1993) leading directly to the
hierarchical configuration now observed.  The cloud core collapsed
until it was a few thousand AU in diameter, at which point it
fragmented into three parts.  The two smaller parts ended up as vB~8
and Gliese~643. The larger fragment continued to contract by another
three orders of magnitude, until a hydrostatic core was formed with a
radius of a few AU (e.g., Larson 1969).  When the temperature of the
core reached about 2000 K the molecular hydrogen began to dissociate,
resulting in a second collapse within the hydrostatic core (e.g.\
Larson 1969).  During the second collapse the core may have been able
to fragment again into the components of the visual binary,
Gliese~644A and Gliese~644B, the latter of which fragmented again into
Ba and Bb.

This scenario is not free of some problems. Although fragmentation
calculations have successfully simulated the formation of wide binaries
during the collapse of molecular cloud cores, there is still some
uncertainty about whether or not this step can succeed in forming a
stable binary with separation as small as 1 AU (e.g., Boss 1989, Bonnell
\& Bate 1994, Bate 1998). Moreover, there have not yet been successful
fragmentation calculations for the formation of very close binaries,
such as Gliese 644B with a separation of only 0.05 AU.  Whatever
mechanism was responsible for the formation of Gliese 644B system, it is
interesting that the mass ratio for this binary is close to unity, a
mass ratio which is apparently quite frequent in spectroscopic binaries
(Tokovinin 2000). A
mechanism that can lead, under the right conditions, to nearly equal
masses in a close binary has been studied by Bate \& Bonnell (1997) and
Bate (2000).  Initially, two stellar seeds form, with masses at most a
few percent of the final stellar masses.  Most of the mass of the
collapsing cloud core is still in a gaseous envelope surrounding the
seeds.  The accretion of material onto the seeds then builds the masses
up to the final stellar values.  The smaller of the two seeds is favored
in the amount of mass that it receives, and this tends to equalize the
masses of the final components of the binary, after the envelope
material has been exhausted.

If the two orbits in the Gliese 644 system are coplanar, as
suggested by our spectroscopic results, then this might lend support to
fragmentation as the formation mechanism for the innermost binary; it
is natural to suppose that the two orbital planes for Gliese~644
should both be oriented perpendicularly to the angular momentum vector
of the molecular cloud core from which they formed. 

Dynamical evolution in a small-N cluster (e.g., McDonald \& Clarke
1993, 1995) is another possible scenario for the formation of the
Gliese~644/643/vB~8 system.  In this scenario all five stars were
formed independently within a loosely self-gravitating small-N cluster
of stars.  Gravitational interactions between the cluster members then
evolved the orbits of the five stars, the ones we see now, into a
hierarchical configuration, while ejecting the other stars from the
cluster.  In this scenario, the more massive stars are expected to
settle into more tightly bound orbits, while the less massive stars
are raised into less tightly bound orbits, just as is observed for the
system.  However, this scenario does not easily account for the
coplanarity of the Gliese~644 triple, a feature often found in other
triple systems as well (Fekel 1981, Tokovinin 1997). 

Another possible way to explain the coplanarity of the close triple
Gliese 644 involves tidal interactions. It is well known that a
large relative inclination between the two orbits of a triple can
induce strong oscillations of the inner eccentricity (e.g., Mazeh,
Krymolowski \& Rosenfeld 1997), which can make the system dynamically
unstable.  Therefore, the low relative inclination of the two orbits
in the Gliese~644 triple might be the result of some selective
evolutionary process. Triple systems with large relative inclinations
simply could not survive. This argument still has to be worked out,
though, because the amplitude of the eccentricity modulation does not
depend linearly on the relative inclination (e.g., Holman, Touma \&
Tremaine 1997), and there are medium-sized angles, on the order of
10--20 degrees, that do not produce large eccentricity modulations. 

A potentially interesting feature of the Gliese 644 triple --- the
very marginal evidence for a minute variation of the outer orbit's
eccentricity --- might also be a result of tidal interactions (Mazeh \&
Shaham 1979, Paper II).  A few spectroscopic triple systems have
already been observed to display direct or indirect evidence for
long-term tidal modulations (Mayor \& Mazeh 1987, Mazeh 1990, Jha et
al.\ 1997), of which HD 109486 is the best example (Jha et al.\ 2000).
Right now the effect observed in Gliese 644 is much too marginal to
rely on.  Many more years of observations will be needed to see if the
effect is real. 

The Gliese 644/643/vB~8 system is the nearest known quintuple stellar
system. The next nearest system with five or more objects is Castor
(=Gliese~278) at 16~pc (Tokovinin 1999).  In order to better understand
how multiple systems of low-mass stars form and evolve, it would be
extremely interesting to find more M-star multiple systems in the solar
neighborhood and to compare their features.  In this vein, it is amazing
that another {\it triple} M-star system exhibits very similar features
to Gliese~644. This is Gliese~866, which was recently studied by
Delfosse et al.\ (1999b) and by Woitas et al.\ (2000). Delfosse et al.\
derived the two orbital periods, of 803 and 3.78 days, and masses of
0.12, 0.11 and 0.096 \msun. Woitas et al.\ (2000) derived a period of
821 days for the outer orbit.  Although the masses in the Gliese~644
triple are almost three times larger, in both systems there are three M
stars with similar masses, where the most massive star is 30\% more
massive than the lightest one, and the period ratio is slightly larger
than 200. Moreover, both systems are probably very close to coplanar.
One wonders if these features are common in M-star triples.

\section*{Acknowledgments}

We thank Jim Peters, Ed Horine, Bob Davis, Dick McCroskey, Skip
Schwartz, Perry Berlind, Marc Payson, Ale Milone, Joe Caruso, Joe Zajac,
Bob Mathieu, Mike Calkins, John Stauffer, and Larry Marschall for making
many of the observations, and Bob Davis for managing the database of CfA
Digital Speedometer observations. We are grateful to the late Charles
E. Worley for providing us with a listing of the astrometric
observations, extracted from the Washington Visual Double Star
Catalog. The referee is thanked for his wise comments and advice. This
research has made use of the SIMBAD database, operated at CDS,
Strasbourg, France.  We gratefully acknowledge support from the
US-Israel Binational Science Foundation grant no. 94-00284 \& 97-00460,
and by the Israel Science Foundation.

\end{document}